\documentclass[fleqn,usenatbib]{mnras}

\usepackage[T1]{fontenc}

\DeclareRobustCommand{\VAN}[3]{#2}
\let\VANthebibliography\thebibliography
\def\thebibliography{\DeclareRobustCommand{\VAN}[3]{##3}\VANthebibliography}

\usepackage{graphicx}	
\usepackage{amsmath}	
\usepackage{amssymb}	
\usepackage{lscape}
\usepackage{ulem}
\usepackage{tabularx}
\usepackage[dvipsnames]{xcolor}
\usepackage{soul}
\usepackage{caption}
\usepackage{subcaption}

\graphicspath{{images/}}

\title[Astro-photometric study of M37]{Astro-photometric study of M37 with \textit{Gaia} and wide-field \textit{ugi}-imaging }

\author[M. Griggio et al.]{M. Griggio$^{1,2}$\thanks{E-mail: massimo.griggio@inaf.it}, L. R. Bedin$^{2}$, 
R. Raddi$^{3}$,
N. Reindl$^{4}$,
L. Tomasella$^{2}$,
M. Scalco$^{1,2}$,
M. Salaris$^{5,6}$,
\newauthor
S. Cassisi$^{6}$,
P. Ochner$^{2,7}$,
S. Ciroi$^{7}$,
P. Rosati$^{1}$,
D. Nardiello$^{2}$,
J. Anderson$^{8}$, 
M. Libralato$^{9}$,
\newauthor
A. Bellini$^{8}$,
A. Vallenari$^{2}$,
L. Spina$^{2}$ and
M. Pedani$^{10}$
\\
$^{1}$Dipartimento di Fisica, Universit\`a di Ferrara, Via Giuseppe Saragat 1, Ferrara I-44122, Italy\\
$^{2}$INAF - Osservatorio Astronomico di Padova, Vicolo dell'Osservatorio 5, Padova I-35122, Italy\\
$^{3}$Universitat Polit\`ecnica de Catalunya, Departament de F\'isica, c/ Esteve Terrades 5, 08860, Castelldefels, Spain\\
$^{4}$Institut f\"ur Physik und Astronomie, Universit\"at Potsdam, Karl-Liebknecht-Stra{\ss}e 24/25, 14476, Potsdam, Germany\\
$^{5}$Astrophysics Research Institute, Liverpool John Moores University, 146 Brownlow Hill, Liverpool L3 5RF, UK\\
$^{6}$INAF - Osservatorio Astronomico di Abruzzo, Via M. Maggini, I-64100 Teramo, Italy\\
$^{7}$Department of Physics and Astronomy, University of Padova, Via F. Marzolo 8, I-35131 Padova, Italy\\
$^{8}$Space Telescope Science Institute, 3700 San Martin Drive, Baltimore, MD 21218, USA\\ 
$^{9}$AURA for the European Space Agency (ESA), Space Telescope Science Institute, 3700 San Martin Drive, Baltimore, MD 21218, USA\\
$^{10}$Fundaci\'on Galileo Galilei - INAF, Rambla Jos\'e Ana Fernandez P\'erez 7, E-38712 Bre\~na Baja (TF), Spain
}

\date{Accepted 2022 July 6. Received 2022 July 5; in original form 2022 June 15}

\pubyear{2022}

\usepackage{newtxtext,newtxmath}
\begin{document}
\label{firstpage}
\pagerange{\pageref{firstpage}--\pageref{lastpage}}
\maketitle

\begin{abstract}
We present an astrometric and photometric wide-field study of the Galactic open star cluster M37 (NGC\,2099). 
The studied field was observed with ground-based images covering a region 
of about four square degrees in the {\it Sloan}-like filters $ugi$. 
We exploited the {\it Gaia} catalogue to calibrate the geometric distortion of the large field mosaics, developing software routines that can be also applied to other wide-field instruments.
The data are used to identify the hottest white dwarf (WD) member candidates of M37.
Thanks to the {\it Gaia} EDR3 exquisite astrometry we identified seven such WD candidates, one of which, besides being a high-probability astrometric 
member, is the putative central star of a planetary nebula. 
To our knowledge, this is a unique object in an open cluster, and we have obtained follow-up low-resolution spectra that are used for a qualitative characterisation of this young WD.
Finally, we publicly release a three-colour atlas and a catalogue of the sources in the field of view, which represents a complement of existing material.
\end{abstract}




\section{Introduction}

The vast majority of low- to intermediate-mass stars in the Galaxy end their lives as white dwarfs (WDs). Their nature of compact objects has served as an important test case for many areas of fundamental physics and stellar evolution theories. WDs in open clusters (OCs) represent a unique opportunity to study these objects in well characterised environments, as OCs usually have very well determined properties such as age, distance, metallicity and reddening.

M37, also known as NGC\,2099, is a rich \citep[with an estimated total mass of $\sim 1500$\,M$_\odot$ by][]{2008A&A...477..165P}, intermediate-age OC, with an angular size of about 1\,deg \citep{2022MNRAS.511.4702G}. As revealed by \cite{2018ApJ...869..139C}, this cluster exhibits an extended Main-Sequence (MS) Turn-Off, thus its precise age is still under debate; \cite{mermilliod1996red} gives an age of about 400-500\,Myr, compatible with the more recent estimate by \cite{2018ApJ...869..139C} that gives an age of 550\,Myr. M37 is located in the Auriga constellation, at a distance of about $1.4$\,kpc \citep[][]{2005A&A...438.1163K,2022MNRAS.511.4702G}. It has been the object of several photometric studies \citep[e.g.,][]{kalirai01} as well as spectroscopic investigations \citep[e.g.,][]{Pancino_2010}. Both photometric and spectroscopic analyses mostly agree on the metallicity of this cluster, which is around solar ($\rm [Fe/H]=0.02$-$0.08$, \citealt{2019MNRAS.490.1821C,2016A&A...585A.150N,2014A&A...561A..93H}), and give a reddening estimate in the range 0.2-0.3 \citep[e.g.,][]{2004AJ....127..991S,Kang_2007}.
M37 has been long known to host quite a large population of WD candidates, consisting of $\sim$~50 stars that were identified via deep $B$- and $V$-band photometry down to $V = 23.5$\,mag \citep{kalirai01} and were characterised through optical spectroscopy \citep{kalirai2005}. More recently, spectroscopic follow-up have confirmed, rejected, or identified new cluster members \citep{2015ApJ...807...90C}, including a very massive $\sim 1.28$\,M$_\odot$ object \citep{cummings2016}. 
The WD census of this OC is not yet complete due to several reasons, like source crowding, dispersal of cluster members, unresolved binarity with MS companions; current estimates outnumber the WD candidates by a factor of 2-4 \citep{kalirai01,richer2021}. Despite their faintness, 10-15 confirmed WD members have been extensively studied for the characterisation of the initial-to-final-mass relation (IFMR), leading to an increasingly improved understanding of mass loss for low- to intermediate-mass stars \citep[][]{ferrario2005,catalan2008,salaris2009,cummings2018}.
\\

In this paper we employ observations of M37 collected at the Asiago Schmidt telescope\footnote{\url{https://www.oapd.inaf.it/asiago/scientific-information-about-telescopes-research/telescopes-and-instrumentations/schmidt-6792}} in the {\it Sloan}-like $ugi$ filters to develop and test a procedure to calibrate the geometric distortion of the instrument, exploiting the {\it Gaia} Early Data Release\,3 \citep[EDR3,][]{2021A&A...649A...1G} absolute reference system, which will be applied also to other wide-field imager mosaics.

By using the method of astrometric cluster membership presented in \cite{2022MNRAS.511.4702G}, we have identified seven new WD member candidates that are currently at the hot end of the WD cooling sequence, which has so far been overlooked by previous searches. These new candidates include the likely central star of a planetary nebula, for which we have also obtained two sets of follow-up spectra. 
We analyse their physical properties and further discuss their cluster membership.

In Sections \ref{sec:data} and \ref{sec:red} we describe the observations and the data reduction routines.
In Sections \ref{sec:memb} and \ref{sec:cmd} we use the membership probability and the astrometric parameters to select cluster members, and we show the colour-magnitude diagrams in the $ugi$ filters.
The newly identified WD member candidates and the their spectral energy distribution analysis are presented in Section \ref{sec:wds}.
Section \ref{sec:cat} describes the catalogue that we publicly release.


\section{Data set}
\label{sec:data}

\begin{figure}
    \centering
    \includegraphics[width=\columnwidth]{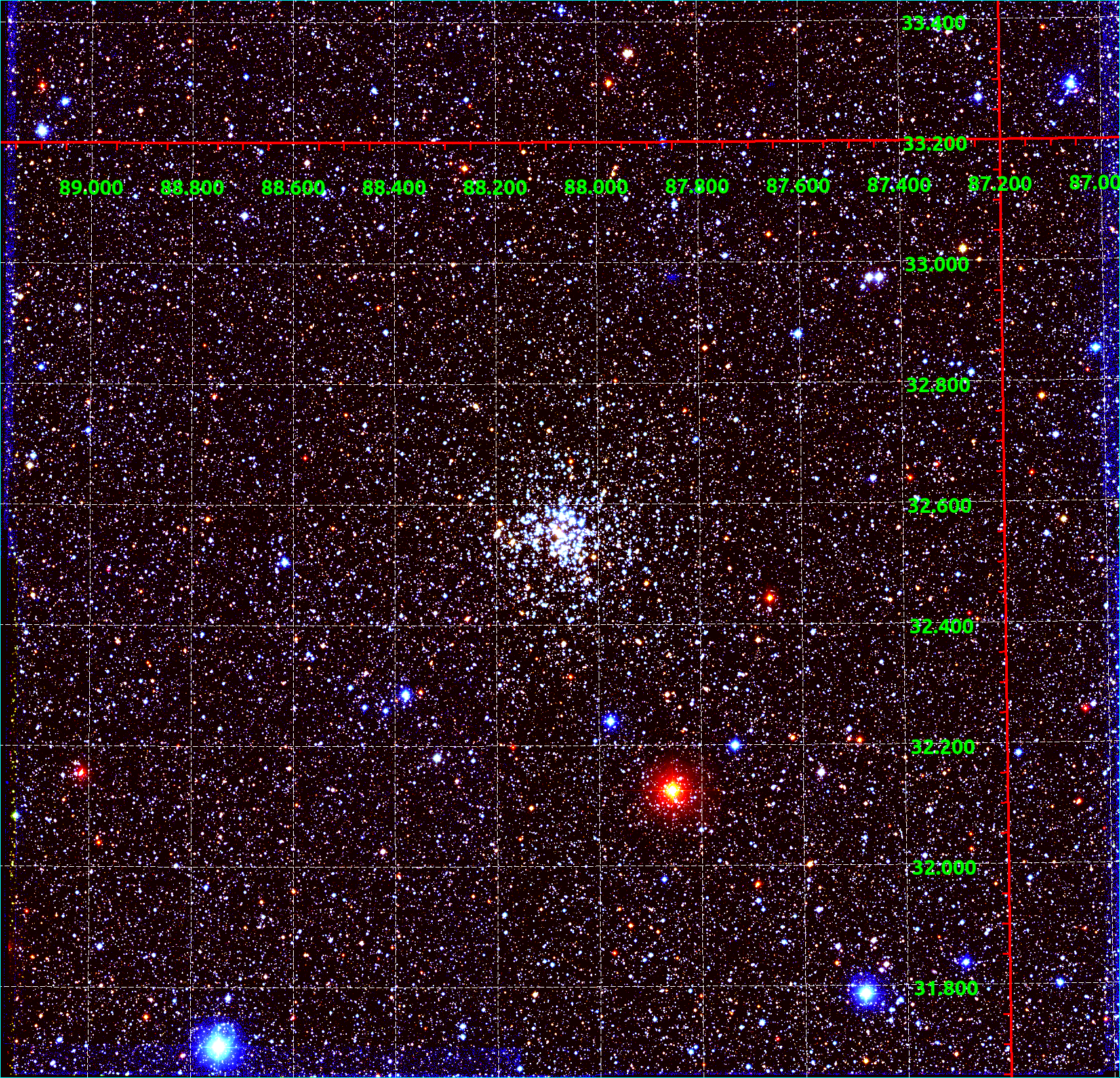}
    \caption{Stacked three-colour image (u,g,i) of the field of view, with superimposed a grid with equatorial ICRS coordinates.}
    \label{fig:stack_rgb}
\end{figure}

The data were collected with the Schmidt $67/92$\,cm telescope in Asiago (Italy)
between 2020 November, $8^{\rm th}$ and $21^{\rm st}$. The telescope is equipped with a KAF-16803 CCD, with an active area of $4096\times4096$ pixels and a field of view (FOV) of $59\times59$\,arcmin$^2$,
that corresponds to a $\sim 0.87$\,arcsec/px scale.
We collected a mosaic of about 2$\times$2\,deg$^2$, with a $\sim 13$ arcmin overlap. 
All the tiles of the mosaic were dithered, and images collected in three filters, $u$-,$g$-, $i$-{\it Sloan}, for a total of 198 images (some of which were later discarded due to poor image quality) each with a $240$\,s exposure time. 
In addition, we collected a set of images with a $10$\,s exposure time to have a better estimation of the flux of very bright sources that are saturated in the $240$\,s data. Table \ref{tab:exposures} summarises the observation log.

\begin{table}
\centering
\caption{Data set used in this work. All the observations were carried out in November 2020.}
\begin{tabularx}{\columnwidth}{lXXX}
\hline
\hline
Filter & \# of exp. $ \times \, t_{\mathrm{exp}}$ & Airmass (arcsec)     & Seeing (arcsec)         \\ 
       &                                          &         (best-worst) &        (best-worst)     \\ \hline
$i$      & $66\times240$\,s & 1.03-1.72 & 1.51-3.22 \\
         & $22\times10$\,s  & 1.03-1.73 & 1.47-3.04 \\ \hline
$g$      & $66\times240$\,s & 1.03-1.84 & 1.53-3.25 \\
         & $22\times10$\,s  & 1.03-1.85 & 1.52-3.05 \\ \hline
$u$      & $66\times240$\,s & 1.04-1.97 & 1.55-2.40 \\
         & $23\times10$\,s  & 1.04-1.98 & 1.53-2.63 \\ \hline
\end{tabularx}
\label{tab:exposures}
\end{table}

The mosaic covers an area centred on M37. A stacked image in three-colour version is shown in Figure \ref{fig:stack_rgb}; a description on how the stack was produced will be given in Section \ref{sec:2p}.

\section{Data reduction}
\label{sec:red}

The data have been corrected via standard calibrations (bias, dark, linearity and flat field). In addition, we accounted for the effects of geometric distortion on the pixels' area in different positions of the CCD, which result into wrong estimations of the flux. In particular, for each pixel we computed a correction factor as the ratio of the area of the distortion-corrected pixel (see \ref{gd_schm}) and the area of the raw pixel (by definition equal to unity).

\subsection{Preliminary photometry}

The first step was to perform a ``preliminary'' photometry (i.e. we extracted only the flux of the brighter stars down to a magnitude limit of $g \simeq 17$), which is then used to compute the transformations between the frames. To this extent we started by using a version of the software by \cite{2006A&A...454.1029A} adapted to the Schmidt's images to derive a grid of empirical point spread functions (PSFs) for each image.
To take into account the spatial variation of the PSF across the FOV, the image is divided into 9$\times$9 regions, and for each region, a PSF is empirically computed using bright, unsaturated and isolated stars. In this way, for each point on the image, a local empirical PSF can be derived by the bilinear interpolation of the 4 closest PSFs.

The grid of PSFs and the image are taken as inputs from a software \citep[described in][and adapted to our detector]{2006A&A...454.1029A} that finds and measures positions and fluxes of the sources in the image by using the local PSF. The routine goes through several iterations progressively finding and measuring fainter sources, until it reaches a threshold limit of $5\sigma$ above the sky background noise. The software outputs a catalogue (one for each image) containing positions and instrumental magnitudes of the stars.

\subsection{Geometric distortion}
\label{gd_schm}

\begin{figure*}
    \centering
    \includegraphics[width=\textwidth]{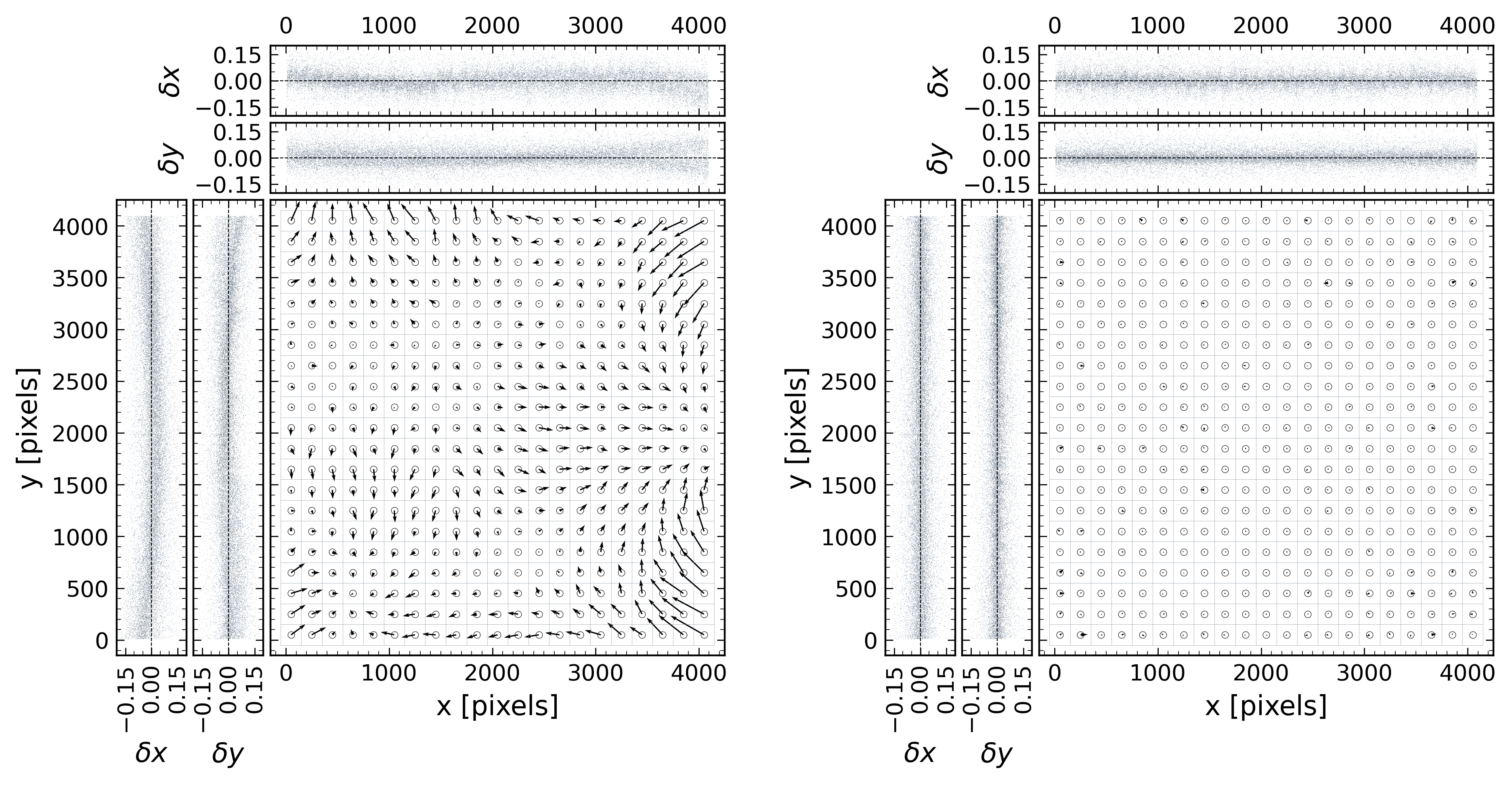}
    \caption{Distortion map for image \texttt{SC132788} before (left) and after (right) the distortion correction routine. The central panels represent the CCD coordinate system, with the arrows showing the magnitude and direction of the distortion on each point, corresponding to a $200\times200$\,pixels box. The arrows have been multiplied by a factor 2500 for visualisation purposes. On the top and left panels we show a projection on the axes of the $x$ and $y$ components of the distortion.}
    \label{fig:gd_sch}
\end{figure*}

To correct for field distortion, we exploited the {\it Gaia} EDR3 absolute reference frame. We cross-referenced the stars in our preliminary catalogues with the stars in the {\it Gaia} EDR3. We used bright, unsaturated sources to derive a general six parameter transformation between the reference frame of our catalogues to the {\it Gaia} reference frame (projected onto the tangent plane of each exposure), and then we computed for each star in each catalogue the residuals between the transformed positions and those given by {\it Gaia}. We divided the residuals from all the catalogues in bins of $200\times200$\,pixels, and for each bin we calculated the mean residual in x and y. The distortion correction routine uses this map of residuals to correct the position of each star via a bi-linear interpolation between the four nearest grid points. In Figure \ref{fig:gd_sch} we show an example of distortion map before and after the correction, left and right, respectively. In the figure we see the CCD divided into $21\times21$ bins, where the arrows represent the mean residual of each bin. Before applying the correction, in particular near the edges of the detector, the distortion goes up to 0.15 pixels, while after the correction have been applied the distortion is less than 0.05 pixels (about 40\,mas).

\subsection{The master frame}
\label{master_frame}

The next step is to put all our images into a common reference frame. This allows us to measure the same star in different exposures simultaneously, thus increasing the signal for fainter sources. The coordinate system on the CCD lies on the plane tangent to the celestial sphere at the central point of the FOV. Since our images have large dithers, each of the $2\times2$ fields lies in a significantly different tangent plane, as discussed also in \cite{2015MNRAS.450.1664L} for a different detector.

In order to derive the coordinate transformations to put every image in the same tangent plane, we took advantage of the {\it Gaia} absolute reference frame. We started by cross-referencing the stars in our preliminary photometry catalogues with the {\it Gaia} EDR3, after correcting their positions for geometric distortions of the field. The {\it Gaia} coordinates were corrected for proper motions to be reported at the epoch of the observations, and then projected onto the tangent plane of each exposure using the procedure described in \cite{2018MNRAS.481.5339B}. This allowed us to derive (for each frame) a global six-parameter transformation from the $(X,Y)$ reference system of the CCD to the meta-coordinate system $(\xi,\eta)$ of its tangent plane. 

We then applied the inverse transformation to project back on the celestial sphere our positions, to have all our preliminary catalogues in spherical coordinates $(\alpha,\delta)$. We then arbitrarily chose a point $(\alpha_0,\delta_0)$ (the same for all the exposures) to project down again all the catalogues onto a common reference system (that now lies in the same tangent plane for every image): in particular we projected all the catalogues on the tangent plane of the image \texttt{SC132969} (archival name), and then applied the inverse transformation from the $(\xi,\eta)$ coordinate system of the tangent plane to the $(X,Y)$ system of the chosen catalogue.

After these steps we have all the catalogues lying in the same plane, in physical pixel coordinates $(X,Y)$, with a mean pixel scale of $868\,$mas/px. For the sake of conciseness, in the following we will call ``preliminary catalogues'' the catalogues after the transformations to get them onto the same coordinate system.

To define the master frame we downloaded a portion of the {\it Gaia} EDR3 catalogue centred on M37 with a radius of $1.5$ deg, we reported the positions of its stars to the observed epoch, we projected it down on the same tangent plane chosen in the previous step and we converted the meta-coordinates to pixel-coordinates by dividing by the pixel scale.
For each filter ($u,\,g,\,i)$ we compiled an average catalogue combining all the single preliminary catalogues and produced a list of the sources found in at least eight images, giving for each source the $3\sigma$-clipped averaged position and magnitude. 
These averaged positions and magnitudes are then cross-referenced with the {\it Gaia} catalogue to obtain a final list of objects with {\it Gaia} positions and $ugi$-magnitudes. This list is then used as master frame: we compute the transformations between each preliminary catalogue to this reference system, which are then used as input in the next step.

\subsection{Second-pass photometry}
\label{sec:2p}
The second-pass photometry is performed by the software \texttt{KS2} \citep[an evolution of the code presented in][]{2008AJ....135.2055A} originally developed for HST images, modified to suit the Schmidt sensor and data taken with large dither patterns. 
Below we give a brief overview of the \texttt{KS2} software, which is described in detail by \cite{Bellini_2017} and \cite{2018MNRAS.481.3382N} (see also \citealt{2021MNRAS.505.3549S} for a more recent application). 

The inputs of the \texttt{KS2} routine are: the images, the PSFs and the transformations derived from the preliminary photometry to find and measure the sources in all the exposures simultaneously. The star finding process goes through different iterations, moving progressively from the brightest to the faintest sources. The software takes as input also a list of bright stars from the preliminary photometry and construct weighted masks around them, which help to avoid PSF-related artefacts. In each iteration the program finds and measures the stars which fit the conditions specified for that iteration, and then subtracts them and proceeds with the next iteration.

\texttt{KS2} measures, in each image, the flux and position of the source using the appropriate local PSF. The star position and flux is determined by fitting the PSF to its $5\times5$ central pixels as in the preliminary photometry, but it computes the final position ad flux as an average between all the images, with the local sky value is computed from the surrounding pixels.

Saturated stars are not measured by \texttt{KS2}; their position and fluxes are recovered by the preliminary photometry and supplemented in the output. To properly measure the stars that were saturated in the $240$\,s exposures we used the $10$\,s images. We performed a second-pass photometry on the short exposures separately, with the zero point registered on the long exposures, and we cross-referenced the sources in the long and short catalogues. We then replaced the magnitudes of the saturated stars in the $240$\,s catalogue with those measured on the $10$\,s frames.

The major upgrade in the code is devoted to perform the projections illustrated in Sec. \ref{master_frame} on the images, to make the code suited to wide-field images 
taken with large dither patterns.
The upgraded code takes as input a set of additional files that are needed to perform the transformations between the local and the master frames. The additional files are:
\begin{enumerate}
    \item one file per image containing the coefficients for the six parameters 
    of the most general linear transformation from the local $(X,Y)$ to the $(\xi,\eta)$ frame of the tangent plane;
    \item one file per image containing the tangent point to each frame used for the projection on spherical coordinates;
    \item one file that contains the $(\alpha_0,\delta_0)$ point needed to project all the frames onto the same tangent plane;
    \item one file that contains the parameters to transform the projected positions on the $(X,Y)$ frame of the chosen image.
\end{enumerate}

The \texttt{KS2} software also outputs an image stack per filter that can be combined for display (as in Figure \ref{fig:stack_rgb}), although they are not suitable for extracting photometry \citep[details on how the stack is produced are given in][]{2008AJ....135.2114A}.
As part of this work, we also made publicly available these atlases of the astronomical scene in the three bands $ugi$.  

\subsection{Input-output corrections}
\begin{figure}
    \centering
    \includegraphics[width=\columnwidth]{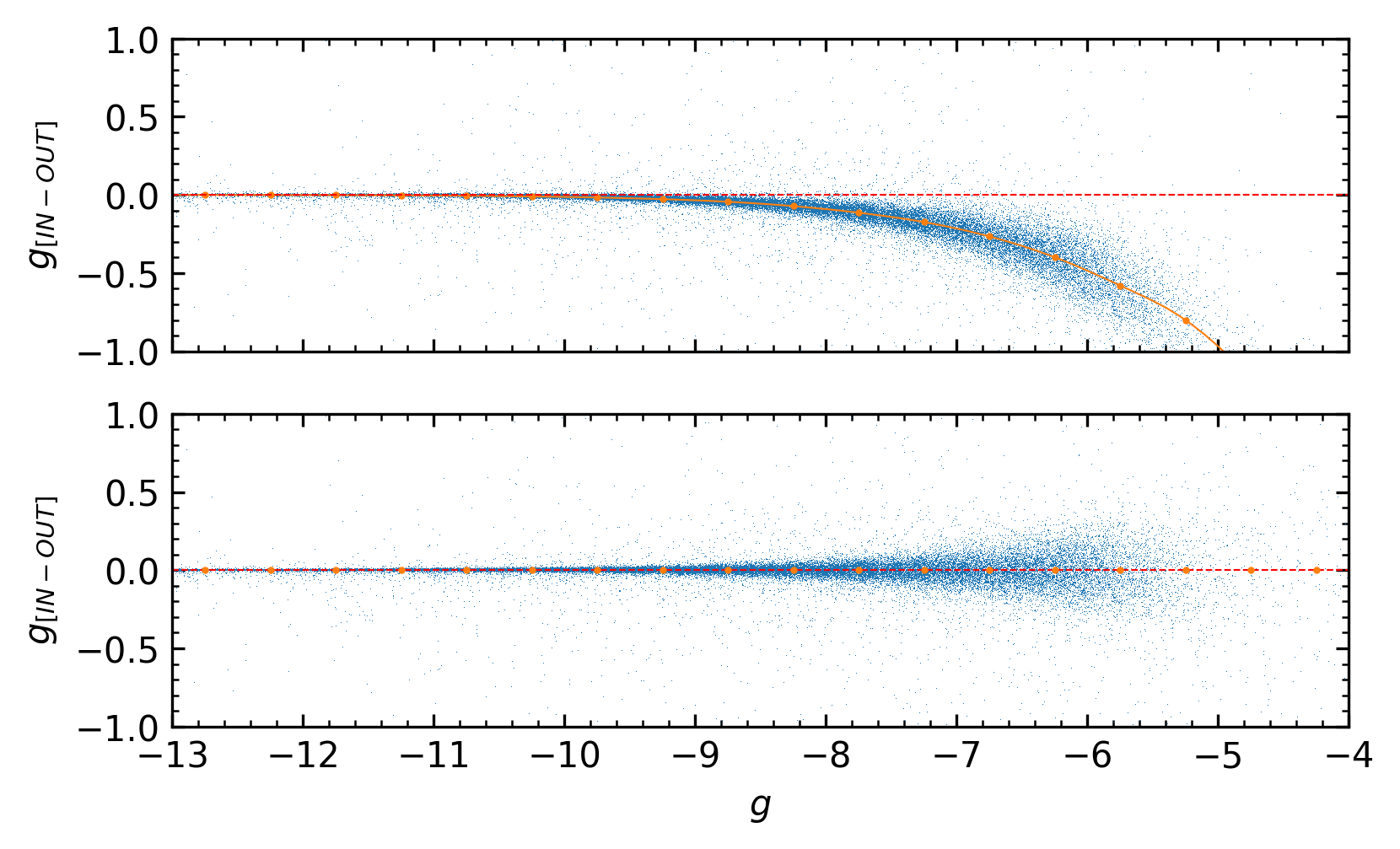}
    \caption{Correction of stellar migration for $g$-filter magnitudes. Top panel: difference between the inserted and the recovered magnitude. Orange dots are the mean residuals for each magnitude bin. The orange line is the interpolating quadratic spline. The dashed line indicates the null difference. Bottom panel: residuals after the correction. Orange dots are the mean residual for each magnitude bin. The dashed line indicates the null difference. Magnitudes on the $x$ axis are instrumental.}
    \label{fig:as}
\end{figure}

In extracting photometry for faint stars, the flux is systematically overestimated when the central pixel of a star coincides with a local maximum of the background noise. This is a well known effect (sometimes referred to as ``stellar migration'') that needs to be corrected for (see \citealt{2009ApJ...697..965B} for details). 
In order to asses the systematic errors, we performed an artificial-star (AS) test using the same procedure described in \cite{2008AJ....135.2055A}. For each AS we chose a random position and $u$ magnitude; the $g-i$ colour is then chosen such that the star falls on the MS ridge-line (drawn by eye). The AS is added in each exposure at the specified position, in the form of an appropriately scaled PSF with Poisson noise. The software routines then operate blindly, finding and measuring all the stars. Examining the output we can determine the accuracy at each magnitude in recovering the AS. Figure \ref{fig:as} (top panel) shows the difference between the inserted and recovered magnitudes for ASs in the $g$ filter, as function of the instrumental magnitude.

To account for these errors we derived a correction in the following way: we started dividing the ASs in bins of $0.5$ magnitudes. We sigma-clipped each bin (at $2\sigma$) around its median to remove outliers, and calculated the median as an estimator of the residual for that bin. We then interpolated the points with a quadratic spline as a function of the magnitude. We used this relation $m_{\mathrm{IN}}-m_{\mathrm{OUT}}$ versus $m$ to correct the effects of the stellar migration on the magnitudes of the stars in our catalogue. In Figure \ref{fig:as} (bottom panel) we show the result of this procedure.

\subsection{Photometric calibration}
\begin{figure}
    \centering
    \includegraphics[width=\columnwidth]{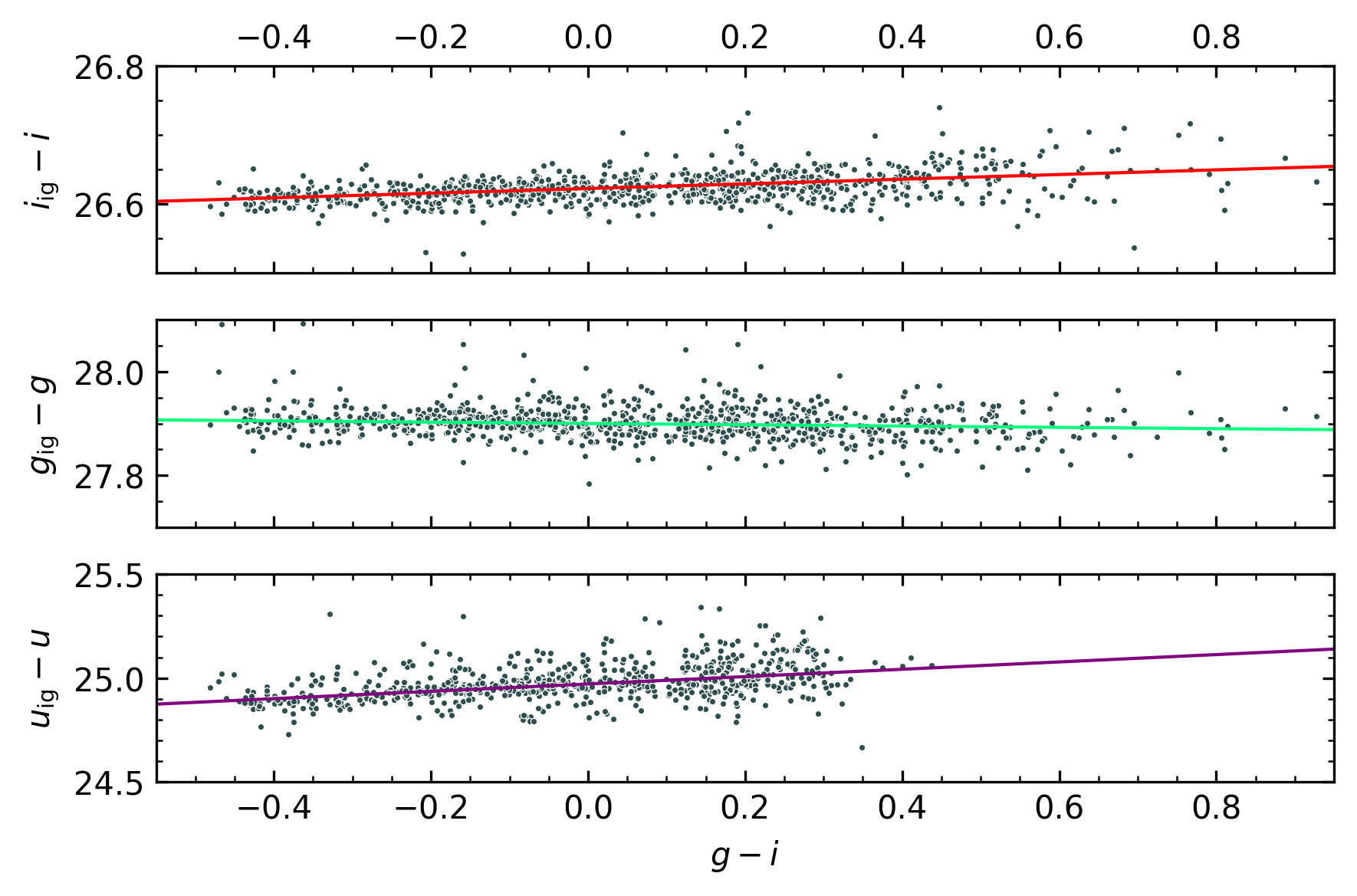}
    \caption{Calibration of the three filters $(i,\,g,\,u)$: the coloured lines represent the linear fit.}
    \label{fig:cal}
\end{figure}
Our instrumental magnitudes have been transformed into the Isaac Newton Telescope (INT) Galactic Plane Survey  \citep[IGAPS;][]{2020A&A...638A..18M} photometric system, that merges the INT Photometric H$\alpha$ Survey \citep[IPHAS;][]{2005MNRAS.362..753D} and the UV-Excess Survey \citep[UVEX;][]{2009MNRAS.399..323G}. We chose the IGAPS catalogue as a reference because, to our knowledge, it is the only one covering the M37 FOV in our three $(u,\,g,\,i)$ filters, though its pass-bands are not exactly equal to ours.
The IGAPS $gri$ photometry is itself uniformly calibrated against the Pan-STARRS system \citep[][]{ps1} with an internal accuracy of 0.02\,mag \citep[][]{2020A&A...638A..18M}. The IGAPS $U_{\rm RGO}$ band is calibrated on a run-by-run basis across the M37 region. 

We calibrated our photometry by using the relation $m_{\rm IGAPS}-m_{\rm instr}$ versus the colour index ($g-i$) in our instrumental magnitudes, as illustrated in Figure \ref{fig:cal}. 
To obtain this relation we cross-matched our sources with the IGAPS catalogue. We considered the stars that were not saturated in both our exposures and IGAPS, and we linearly interpolated (for $m=i,\,g,\,u)$ the $m_{\rm IGAPS}-m_{\rm instr}$ versus $g-i$ distribution. 
A first calibration has been performed using all the unsaturated stars in our catalogue, and after selecting the cluster members (see next section), in order to mitigate potential systematic effects, we further restricted the sample to M37 members only. To transform the instrumental magnitudes we adopted the relation $m_{\rm cal} = m_{\rm instr} + a_0 + a_1 \times (g-i)$, where $m_{\rm cal}$ is our final magnitude calibrated with respect to IGAPS, and the coefficients are determined via a best fit approach.
Higher order polynomials proved to be unnecessary.


\section{Selection of cluster members}
\label{sec:memb}

\begin{figure}
    \centering
    \includegraphics[width=\columnwidth]{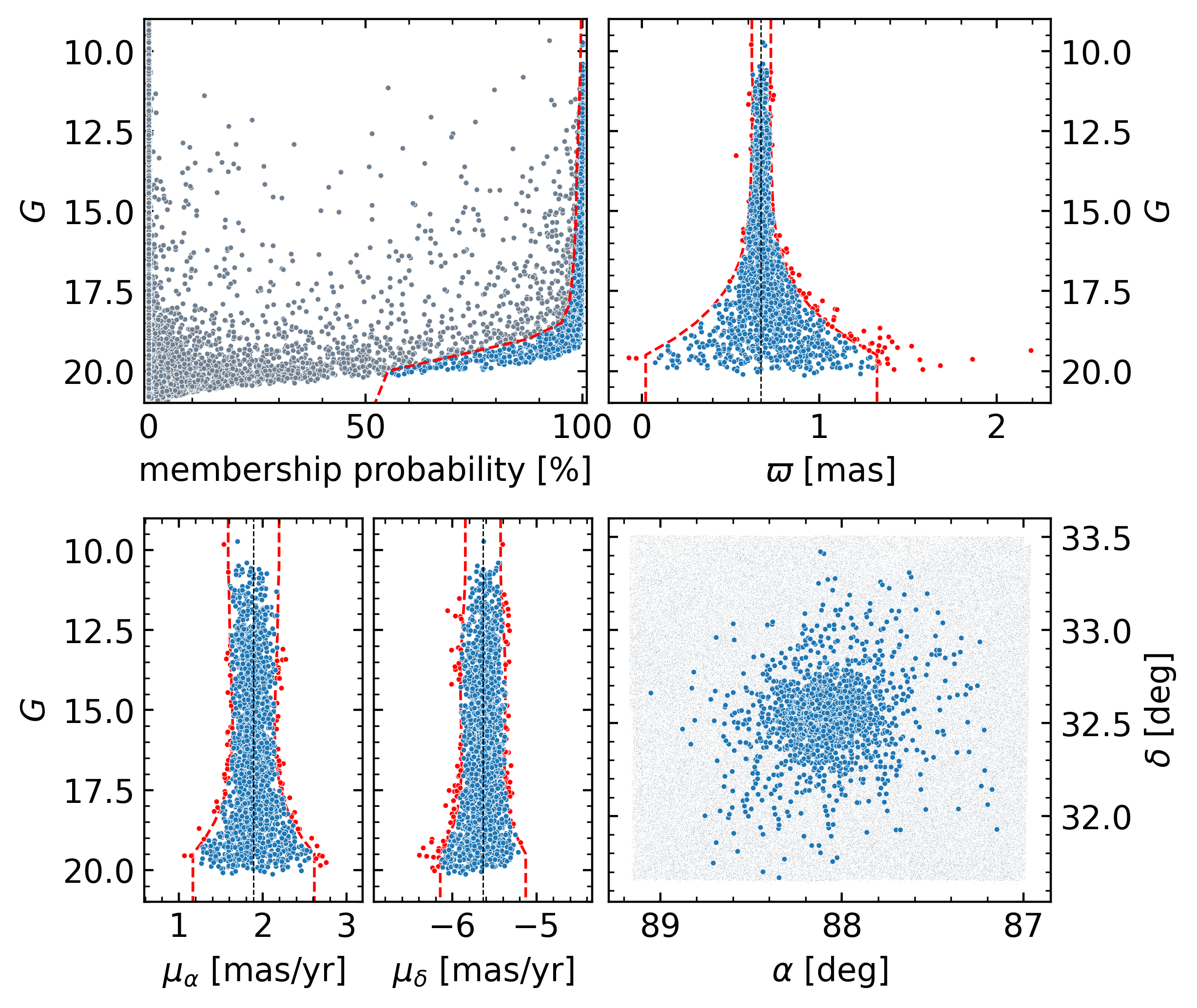}
    \caption{Members selection. Top left: membership probability for all the sources. We reject the stars that fall above the red line. Top right: $G$ magnitude vs parallax. Here we reject the stars that fall outside the region delimited by the two red lines (blue dots). The black dashed line represents the median parallax. Bottom left: proper motions of the sources that passed the previous two cuts vs their $G$ magnitude. We kept the sources between the red dashed lines (blue dots). Bottom right: spatial distribution of the sources. Blue dot are the selected members of M37.}
    \label{fig:memb}
\end{figure}

The selection of member stars have been performed using the membership probabilities derived from \cite{2022MNRAS.511.4702G}.
To select highly-reliable cluster members we adopted the procedure that is displayed in Figure \ref{fig:memb}. In the top left panel, we show the membership probability plotted against the {\it Gaia}-$G$ band magnitude. We applied a by-eye cut 
with the idea of selecting the bulk of sources with cluster membership at each magnitude. 
This cut is indicated by the dashed red curve, and we kept only the sources below that curve. This cut on the membership probabilities becomes less strict going towards fainter magnitudes as the measurement errors increase and members becomes less certain.

This sample has then been constrained on the parallax versus magnitude plane (top right panel). We divided the data into magnitude bins of 1 mag, and for each bin we calculated $\sigma_{\varpi,i}=68.27^{\rm th}$\,percentile of the residuals around cluster's parallax $\varpi$ \citep[given by][]{2022MNRAS.511.4702G}. We derived the red curves interpolating the points $\varpi\pm2\sigma_{\varpi,i}$ with a spline.

The last selection was performed on the proper motions versus $G$ plane. We applied the same procedure as for the parallax to $\mu_\alpha$ and $\mu_\delta$ to derive the red curves.

On the bottom right panel we show the spatial distribution of the stars that are selected as M37 members.


\section{\textit{Sloan} Colour-magnitude diagrams}
\label{sec:cmd}

\begin{figure}
    \centering
    \includegraphics[width=\columnwidth]{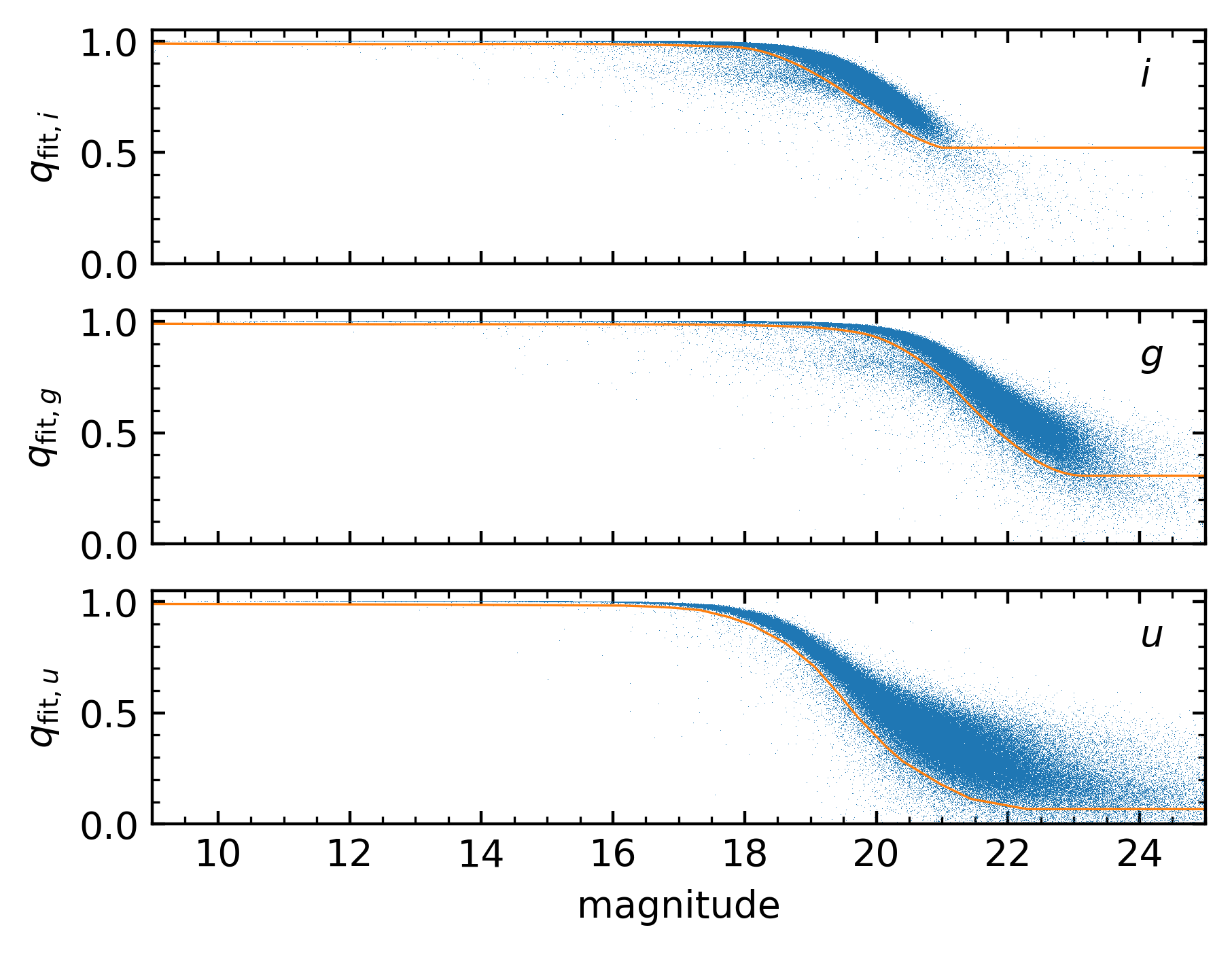}
    \caption{$q_{\mathrm{fit}}$ versus magnitude in each filter. We marked as good (\texttt{pho\_sel\_f}$=1$, $\mathrm{f}=u,\,g,\,i$) the stars that are above the orange line and $3\sigma$ above the sky.}
    \label{fig:pho_sel}
\end{figure}

We first rejected sources with poor measurements, by applying a cut in the quality of fit ($q_{\mathrm{fit}}$) parameter, which is in the \texttt{KS2} output for each source. The $q_{\mathrm{fit}}$ measures how well a source is fitted by the PSF; a $q_{\mathrm{fit}}=1$ indicates a perfect fit. The cuts we performed are shown in Figure \ref{fig:pho_sel}. In addition to these cuts, we selected the sources that are at least $3\sigma$ above the sky background. The sources that passed these selection criteria and are not saturated are flagged with $\texttt{pho\_sel\_f}=1$ in our catalogue, with $\mathrm{f}=u,\,g,\,i$. The additional flag $\texttt{pho\_sel}=1$ is for stars that passed the selection in all the three filters.

\begin{figure*}
    \centering
    \includegraphics[width=\textwidth]{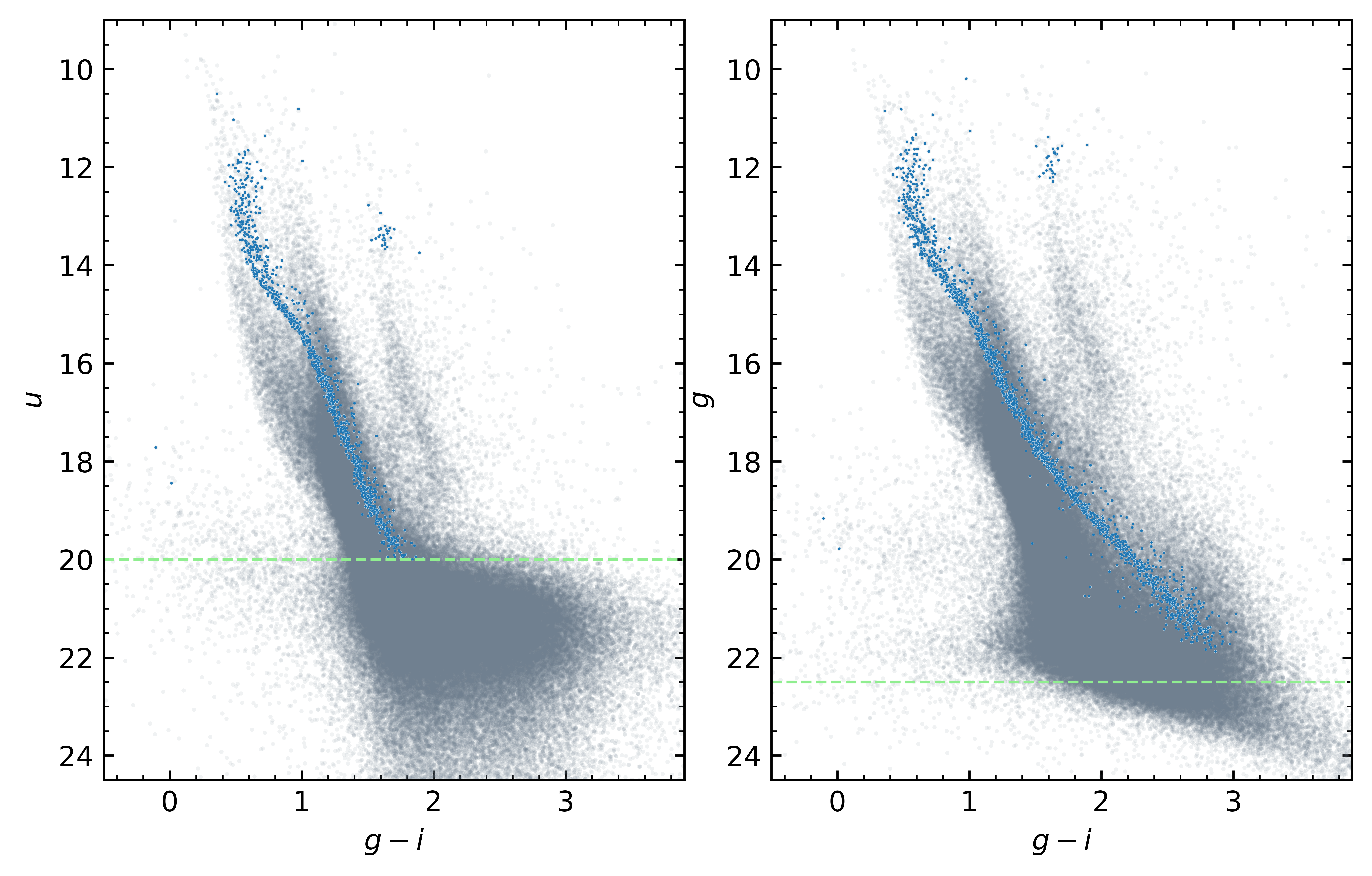}
    \caption{Colour-magnitude diagrams in $ugi$ filters. Grey points represent the sources that passed the quality cuts in the considered filters of each CMD. Blue points are the sources that passed the members selection. The dashed green line represents the $3\sigma$ cut above the sky background level (in $u$ and $g$ respectively).}
    \label{fig:cmd_all}
\end{figure*}

In Figure \ref{fig:cmd_all} we show a colour-magnitude diagram (CMD) of the selected sources in the FOV (in grey) of M37 and the cluster members as defined in previous Section 
(in blue). The dashed green line represents the $3\sigma$ cut above the sky background in the $u$ and $g$ filters.


\section{White Dwarf Member Candidates}
\label{sec:wds}
\begin{figure}
    \centering
    \includegraphics[width=\columnwidth]{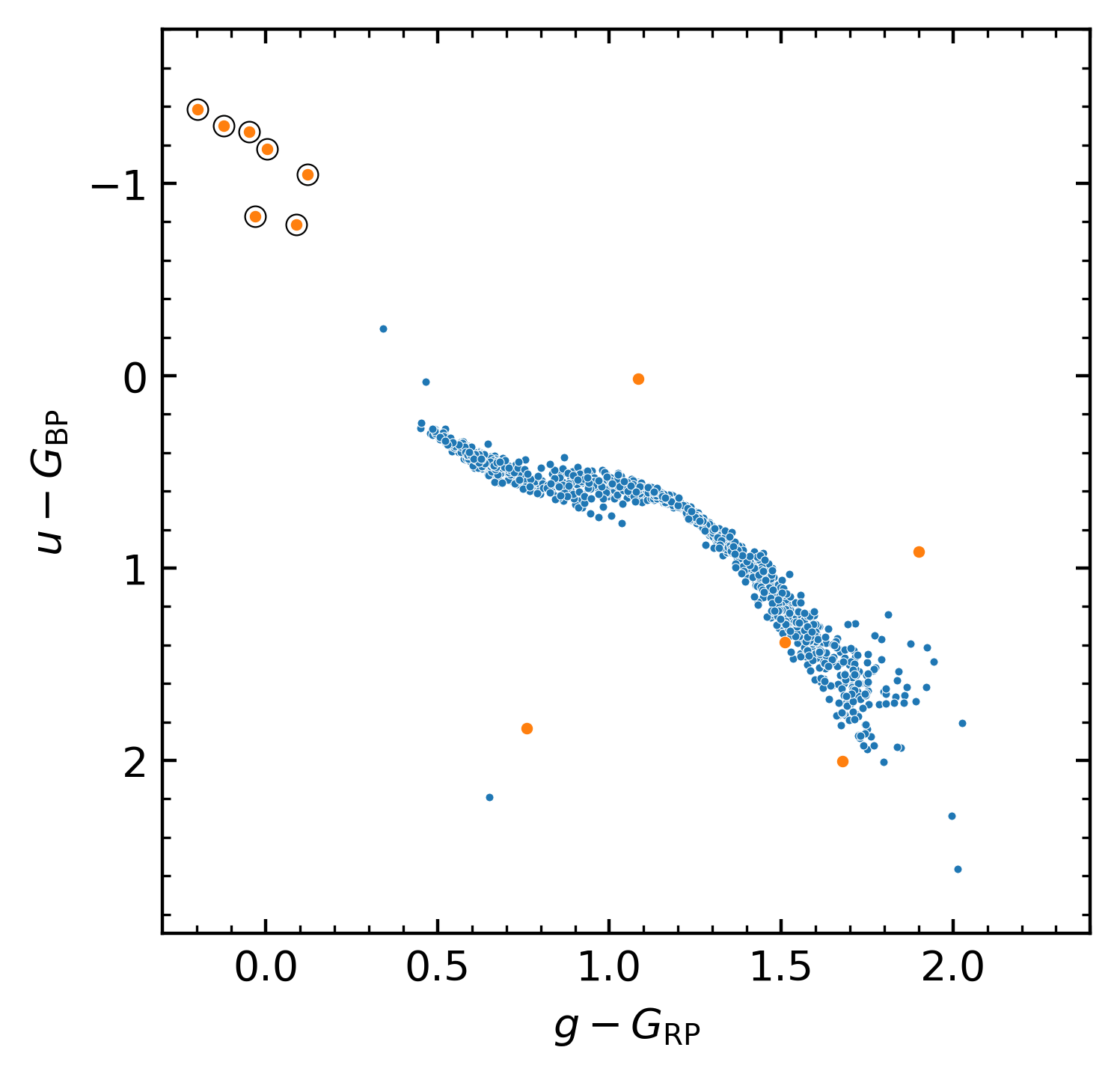}
    \caption{Two colour diagram of member stars. Blue points represent the cluster members. Orange dots indicate the sources that passed the selection described in the text. The orange dots encircled in black are those that we selected as WD member candidates.}
    \label{fig:tcd}
\end{figure}

M37 is known to host a few, faint white dwarf (WD) members, also including a very massive object of $\approx 1.3$\,M$_{\odot}$ \citep{2015ApJ...807...90C,cummings2016}. Given the current age of M37, the lightest WD members that could have formed through canonical single-star evolution are expected to have $\approx 0.7$\,M$_{\odot}$. We focused our search for WD candidates at the hot-end of their cooling sequence that, however, for M37 corresponds to a region of the CMD where the measurement errors of {\it Gaia} are the largest.

We restricted our search to sources 
with $G_{\mathrm{BP}}>18$ and $-0.5<G_{\mathrm{BP}}-G_{\mathrm{RP}}<0.5$ where we expect 
to find WDs that have just recently formed and have both the same distance and reddening of M37 ($d = 1.5\pm 0.1$\,kpc, \citealt{2022MNRAS.511.4702G}; $E(B-V) = 0.26$, \citealt{2018ApJ...869..139C}). We selected from this region the stars with proper motions and parallaxes that are compatible within $3\sigma$ with those of the cluster.
To confirm the isolated WD nature of these objects we built a two-colour $(u-G_{\mathrm{BP}},\, g-G_{\mathrm{RP}})$ diagram (Figure \ref{fig:tcd}), in which we identify the stars that passed this selection with an orange dot. Seven of them possess the typical blue colours of isolated WDs. All the previously known WD members of M37 are fainter than our {\it Gaia}-based selection \citep[][and references therein]{2015ApJ...807...90C,cummings2016}.

\begin{figure*}
    \centering
    \includegraphics[width=\textwidth]{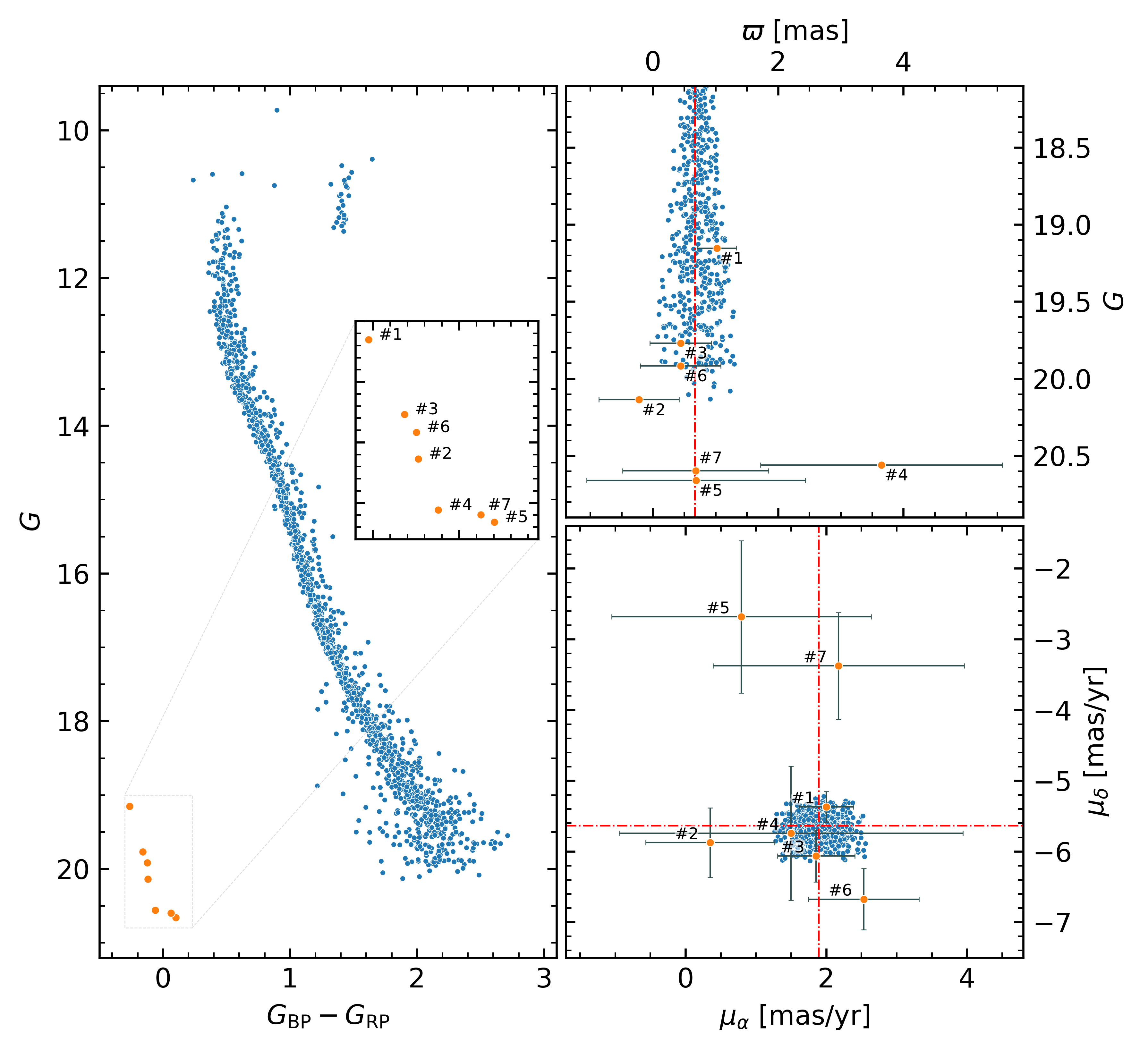}
    \caption{Left: CMD of M37 members in the {\it Gaia} pass-bands (blue). Orange dots are the WD member candidates identified in this work. Error bars for parallax and proper motions are given by {\it Gaia} EDR3 uncertainties. Top right: parallax vs {\it Gaia} $G$ magnitude; the vertical red line represents the parallax of the cluster. Bottom right: Proper motion diagram; the red lines represent the absolute proper motion of the cluster.}
    \label{fig:wd}
\end{figure*}

The astrometric and photometric properties of the seven WD candidates are shown in Figure \ref{fig:wd} along with the other cluster members. Their relevant {\it Gaia} data are listed in Table\,\ref{tab:wd_par}. In Figure \ref{fig:wdfc} we show the finding charts for WD2--7, while that of WD1 is shown in Figure \ref{fig:wd1_spec} (right).

\begin{table*}
    \centering
    \caption{Photometric and astrometric properties  of the seven WD candidates.}
    \begin{tabularx}{\textwidth}{ccccccccccc}
    \hline
    \hline
    \#  & {\it Gaia} EDR3 ID & $G$ & $G_{\rm BP}$ & $G_{\rm RP}$ & $g$ & $u$ & $i$ & $\mu_\alpha$\,[mas/yr] & $\mu_\delta$\,[mas/yr] & $\varpi$\,[mas] \\ \hline
    WD1 & \texttt{3451205783698632704} & 19.154 & 19.100 & 19.362 & 19.164 & 17.716 & 19.271 & $2.003 \pm 0.387$ & $-5.371 \pm 0.219$ & $\phantom{+}1.020 \pm 0.315$   \\
    WD2 & \texttt{3451182182857026048} & 20.137 & 20.038 & 20.156 & 20.160 & 18.859 & 19.923 & $0.353 \pm 0.916$ & $-5.877 \pm 0.490$ & $-0.224\pm 0.640$   \\
    WD3 & \texttt{3451201076423973120} & 19.769 & 19.746 & 19.904 & 19.781 & 18.448 & 19.768 & $1.858 \pm 0.550$ & $-6.064 \pm 0.368$ & $\phantom{+}0.444 \pm 0.492$   \\
    WD4 & \texttt{3451167786125150592} & 20.560 & 20.492 & 20.552 & 20.641 & 19.707 & 20.762 & $1.500 \pm 2.443$ & $-5.743 \pm 0.946$ & $\phantom{+}3.649 \pm 1.932$   \\
    WD5 & \texttt{3451200114340263296} & 20.660 & 20.642 & 20.540 & 20.662 & 19.596 & 20.457 & $0.796 \pm 1.845$ & $-2.684 \pm 1.076$ & $\phantom{+}0.688 \pm 1.746$   \\
    WD6 & \texttt{3448258267902375296} & 19.917 & 19.957 & 20.081 & 20.050 & 19.130 & 20.171 & $2.534 \pm 0.787$ & $-6.675 \pm 0.435$ & $\phantom{+}0.443 \pm 0.643$   \\
    WD7 & \texttt{3454340495645123584} & 20.599 & 20.643 & 20.580 & 20.531 & 19.375 & 20.134 & $2.177 \pm 1.786$ & $-3.378 \pm 0.755$ & $\phantom{+}0.682 \pm 1.164$   \\
    \hline
    \end{tabularx}
    \label{tab:wd_par}
\end{table*}

\subsection{Spectroscopic follow-up of WD1}

We collected optical spectra of the brightest candidate, WD1, with the Asiago Faint Object Spectrograph and Camera (AFOSC) mounted on the 1.82-m Copernico telescope operated by INAF-Osservatorio Astronomico di Padova atop of Mount Ekar, Asiago, Italy. We used the gr4 grating with both 2.5 and 1.69-wide slits, achieving a resolving power of $R = 219$ and $325$ at the central wavelength of gr4 for the two setups, respectively. One single exposure of 3600-s were taken on November 10, 2021, under hazy sky conditions (slit 2.50 arcsec), while three exposures of 2700-s each were taken on January 12, 2022 under good sky conditions (slit 1.69 arcsec). The data were reduced with an \verb|iraf| \citep{iraf} based pipeline, i.e. \verb|FOSCGUI|\footnote{FOSCGUI is a graphic user interface aimed at reducing spectrophotometric data and extracting spectroscopy of the focal reducer type spectrograph/camera FOSC. It was developed by E. Cappellaro. A package description can be found at http://sngroup.oapd.inaf.it/foscgui.html}. 
The average second-epoch spectrum signal-to-noise ratio (S/N) is around 15. A first look of the combined spectrum confirmed this star as a hot, likely hydrogen-deficient star.

\begin{figure*}
    \centering
    \includegraphics[width=88mm]{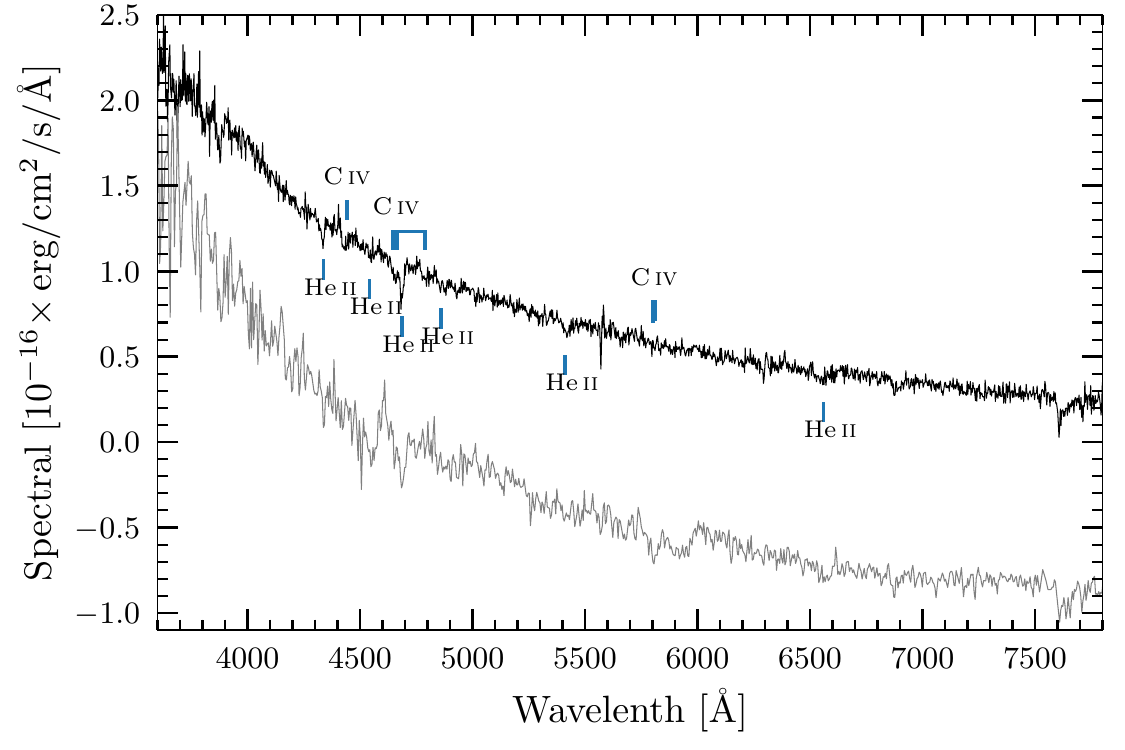} 
    \includegraphics[width=80mm]{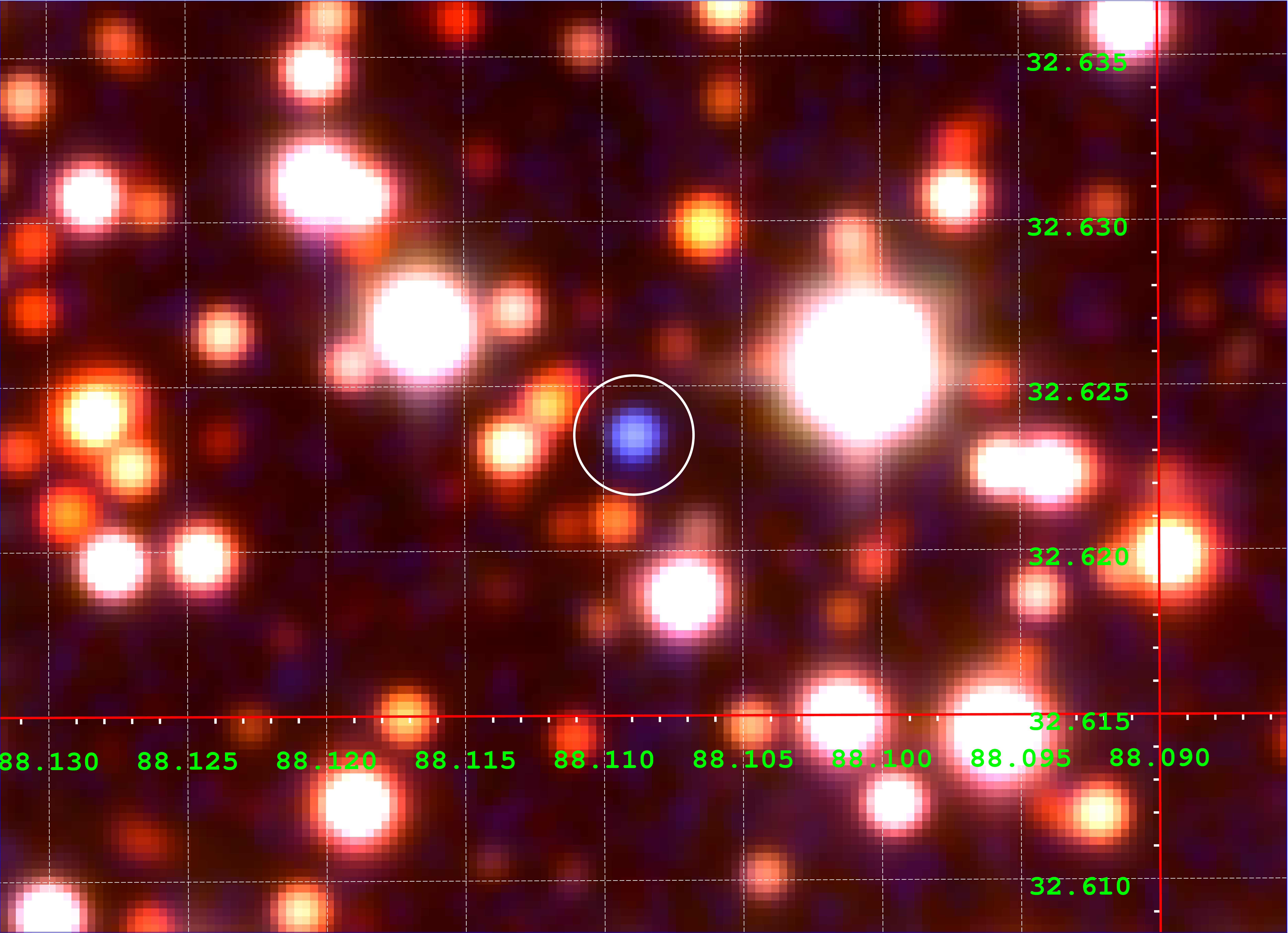} 
    \caption{
Left: The TNG/LRS spectrum (black) and the Asiago/AFOSC spectrum (gray). The latter is shifted vertically by a constant factor. We have labelled some of the most prominent He\,{\sc ii} and C\,{\sc iv} lines. 
Right: Tri-chromatic stack finding-chart of WD1.  
}
    \label{fig:wd1_spec}
\end{figure*}

We re-observed WD1 with the Device Optimized for the LOw RESolution (DOLORES, LRS in short) that is mounted on the Telescopio Nazionale Galileo (TNG) at the Observatorio del Roque de los Muchachos (Canary Islands, Spain).
We took three exposures of 1800 seconds each on 2022-03-03, in service mode during Director Discretionary Time (program A44DDT3). We used the low-resolution blue grating (LR-B) with a 1-arcsec slit that enabled a dispersion of 2.8\,\AA/pixel and a resolution of 11\,\AA\ at 5577\,\AA. The observations were performed in dark time under good weather conditions and with a seeing well below $1$\,arcsec. The spectrophotometric standard Hiltner\,600 was observed at the beginning of the night with the same slit width. The data were reduced, 1D optimally extracted and wavelength and flux calibrated by using standard reduction procedures with the \verb|starlink| \citep{starlink}, \verb|pamela| \citep{pamela}, and \verb|molly| \citep{molly} software packages.
The average LRS spectrum of WD1 has S/N = 35 at 5500\,\AA. In Fig.\,\ref{fig:wd1_spec}, we show both the AFOSC and LRS spectra, labelling the most prominent absorption features of C\,{\sc iv} at $\approx 4660$ and 5801--5812\,\AA\, and the He\,{\sc ii} lines.
\subsection{Physical parameters}
In order to shed more light on the properties of the seven WD candidates, we performed model fits of their Spectral Energy Distributions (SED). We collected available photometry from the {\it Galaxy Evolution Explorer} \citep[GALEX DR6+7;][]{galex},
IGAPS \citep{2020A&A...638A..18M}, PanSTARRS \citep{ps1}, and the {\it Sloan} Digital Sky Survey \citep[SDSS DR12;][]{sdss12}. The SED fitting procedure allowed us to estimate the effective temperature ($T_{\rm eff}$), radius, mass, and cooling age of the candidate WDs 
via $\chi^2$ minimisation of the difference among observed photometry and the appropriate synthetic magnitudes. The latter were computed from a grid of \citet{koester2010} models for hydrogen-dominated WD spectra (DA type), which we mapped on the La Plata cooling tracks \citep{althaus2013,camisassa2016}. We adopted the distance of M37 and the differential reddening estimated \citep[following the procedure described in][]{2017ApJ...842....7B} from the cluster neighbours of each WD candidate as external priors in the SED fitting routine. The results of our SED fitting are shown in Fig.\,\ref{fig:seds} for WD2--WD7. Their photometrically estimated physical parameters are listed in Table\,\ref{tab:wd_phys}. The physical parameters of WD1 that could be obtained from the SED fitting routine would not be reliable, because this star is confirmed not to have a hydrogen-dominated atmosphere (see next section). A spectroscopic follow-up of the most likely candidates (that is WD2, WD3, WD4, and WD5) remains necessary in order to independently confirm their cluster membership.

\begin{table}
\centering
\caption{Physical parameters of six WD candidates obtained from the SED fitting analysis.}
\label{tab:wd_phys}
\begin{tabular}{lrrrr}
\hline
\hline
\# & $T_{\rm eff}$ [K] & $R/\mathrm{R_{\odot}}$ & $M/\mathrm{M_{\odot}}$ & $\tau_{\rm cool}$ [Myr] \\
\hline
\noalign{\smallskip}
  WD2 & $46900_{-14700}^{+20900}$ & $0.019_{-0.003}^{+0.005}$ & $0.52_{-0.10}^{+0.11}$ & $1.9_{-1.5}^{+2.2}$\\
  \noalign{\smallskip}
  WD3 & $64500_{-11700}^{+10000}$ & $0.020_{-0.002}^{+0.002}$ & $0.57_{-0.05}^{+0.05}$ & $0.7_{-0.4}^{+0.8}$\\
  \noalign{\smallskip}
  WD4 & $60800_{-10800}^{+12100}$ & $0.015_{-0.001}^{+0.001}$ & $0.67_{-0.07}^{+0.08}$ & $0.8_{-0.4}^{+1.0}$\\
  \noalign{\smallskip}
  WD5 & $41900_{-9800 }^{+16000}$ & $0.017_{-0.002}^{+0.003}$ & $0.56_{-0.09}^{+0.12}$ & $3.4_{-2.5}^{+3.8}$\\
  \noalign{\smallskip}
  WD6 & $35000_{-2700 }^{+4200}$  & $0.024_{-0.002}^{+0.002}$ & $0.44_{-0.06}^{+0.03}$ & $3.1_{-1.6}^{+1.6}$\\
  \noalign{\smallskip}
  WD7 & $21100_{-2100 }^{+2300}$  & $0.027_{-0.002}^{+0.002}$ & $0.33_{-0.04}^{+0.03}$ & $7.4_{-5.6}^{+20.2}$\\
  \noalign{\smallskip}
\hline
\end{tabular}

\end{table}

\subsubsection{WD1}

\begin{figure*}
    \centering
    \includegraphics[width=\textwidth]{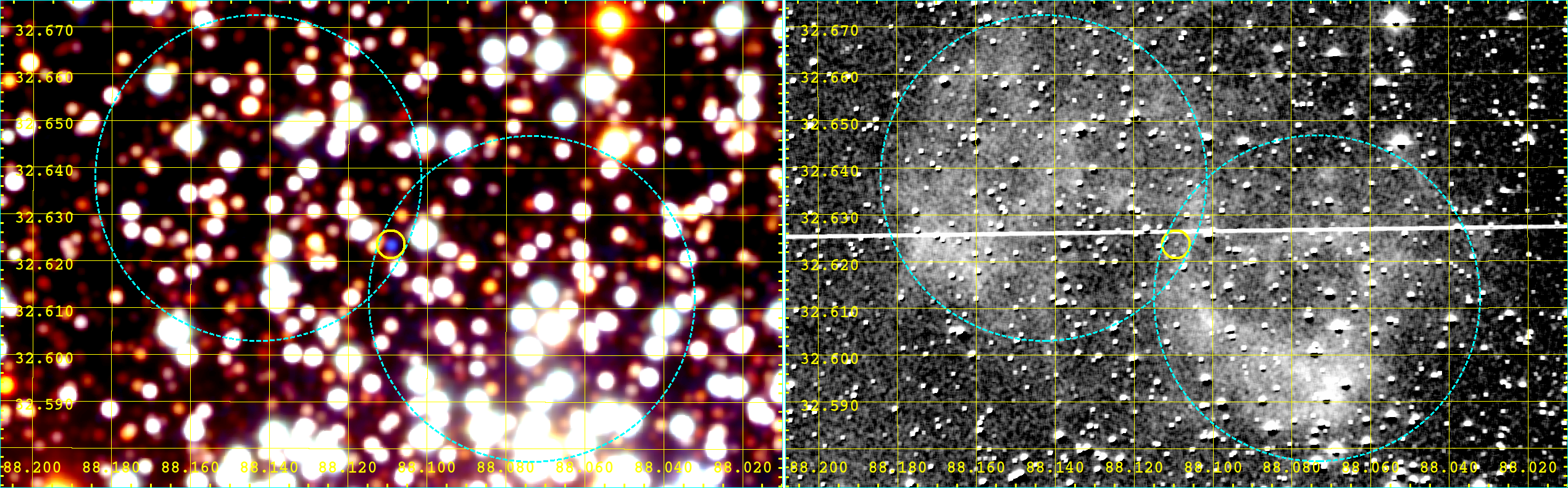}
    \caption{Left: RGB stack of the region around WD1, encircled in yellow. The putative locations of the two lobes of the PN are encircled in cyan. Right: H$\alpha$--$r$ image of the same region, which highlights the nebula, taken from the HASH database \protect\citep[][]{parker2016}.}
    \label{fig:wd1PN}
\end{figure*}
The LRS spectrum of WD1 confirms it as a hot, hydrogen-deficient WD that appears as an intermediate object between the PG\,1159 and DO classes \citep{werner2006,reindl2014}. The spectral resolution and S/N of the spectra in hand are not sufficient for performing a quantitative spectral analysis, but its $T_{\rm eff}$ is likely hotter than 60,000\,K. This star was also identified by \citet{2021A&A...656A.110C} as the central star of a faint, bipolar planetary nebula (PN) \citep[PN\,G177.5+03.1;][]{parker2016}. 
From Fig.\,\ref{fig:wd1PN} a diameter $d = 460$\,arcsec can be estimated (consider $d=4 \times r$, with $r$ being the radius of one lobe). Assuming furthermore a distance of 1.5\,kpc \citep[][]{2022MNRAS.511.4702G}, and an expansion velocity of
20\,km/s, one can estimate the age of the PN to be 82\,kyr. Such an old PNe can be expected for a star that has already entered the WD cooling sequence. 
The relation of this PN (shown in Figure \ref{fig:wd1PN}) with WD1 needs further confirmation: 
additional follow-up observations are needed to better characterise the evolutionary status and cluster membership of WD1 (that appears as a high-probability astrometric member of this cluster, having passed the membership selection criteria), via quantitative spectral analysis. If its physical properties and radial velocities turn out to be also compatible with the expected parameters (a mass that is larger than $0.7$\,M$_\odot$, as well as a total age and mean radial velocity like those of the other M37 members) WD1 would represent a very rare (if not unique) central star of a planetary nebula (CSPN) that has passed all the tests of open cluster membership \citep{moni-bidin2014, fragkou2019, 2020AJ....159..276B}. 
\subsubsection{WD2, WD3 and WD5}
For these three stars we have estimated photometric masses that are within 2 or 3-sigmas from the minimum allowed mass for single star evolution of $0.7$\,M$_{\odot}$. The estimated cooling ages are compatible with the total age of M37. In addition, WD3 passed the membership selection criteria.
\subsubsection{WD4}
This object is the most likely cluster member confirmed at the 1-$\sigma$ level, as we obtain a $T_{\rm eff} = 60\,800^{+12\,100}_{\,-10800}$\,K, a mass of $0.67\pm0.08$\,M$_{\odot}$, and a cooling age of $0.8^{+1.0}_{-0.4}$\,Myr. If confirmed via spectroscopic follow-up, this WD could be one of the youngest evolved members of M37.

\subsubsection{WD6 and WD7}
Although they are likely very young, these two WD candidates have
photometric estimates of radii and masses that are not compatible
with their proposed membership to M37. In fact, masses lower than 0.5\,M$_\odot$
are indicative of He-core WDs, originated from RGB stars with electron 
degenerate helium cores, that have lost their envelope before the onset of the 
helium flash. Given the age of this cluster, the presence of He-degenerate
cores in the post-MS stars is very unlikely (theoretical isochrones predict post-MS stars 
with initial masses larger than 2.5\,M$_\odot$). Therefore, we do not
believe that these two WDs can belong to M37.


\section{A new catalogue}
\label{sec:cat}
\begin{table}
    \centering
    \caption{Description of the columns in our catalogue.}
    \begin{tabularx}{\columnwidth}{lX}
    \hline
    \hline
    \textbf{Column} & \textbf{Description}   \\ \hline
    \texttt{gid} & {\it Gaia} EDR3 id of the source \\
    \texttt{ra} & right ascension [deg] \\
    \texttt{dec} & declination [deg] \\
    \texttt{i} & magnitude in the $i$ filter \\
    \texttt{ei} & error on the $i$ mag \\
    \texttt{g} & magnitude in the $g$ filter \\
    \texttt{eg} & error on the $g$ mag \\
    \texttt{u} & magnitude in the $u$ filter \\
    \texttt{eu} & error on the $u$ mag \\
    \texttt{qi} & quality flag for $i$ mag \\
    \texttt{qg} & quality flag for $g$ mag \\
    \texttt{qu} & quality flag for $u$ mag \\
    \texttt{oi} & fraction of flux within the PSF aperture in $i$ filter due to neighbours \\
    \texttt{og} & fraction of flux within the PSF aperture in $g$ filter due to neighbours \\
    \texttt{ou} & fraction of flux within the PSF aperture in $u$ filter due to neighbours \\
    \texttt{pho\_sel} & 1 if source passed photometric cuts in all filters\\
    \texttt{pho\_sel\_i} & 1 if source passed photometric cuts in the $i$ filter \\
    \texttt{pho\_sel\_g} & 1 if source passed photometric cuts in the $g$ filter \\
    \texttt{pho\_sel\_u} & 1 if source passed photometric cuts in the $u$ filter \\
    \texttt{G} & magnitude in the {\it Gaia} EDR3 $G$ filter \\
    \texttt{eG} & error on the $G$ mag \\
    \texttt{Gbp} & magnitude in the {\it Gaia} EDR3 $G_{\mathrm{BP}}$ filter \\
    \texttt{eGbp} & error on the $G_{\mathrm{BP}}$ mag \\
    \texttt{Grp} & magnitude in the {\it Gaia} EDR3 $G_{\mathrm{RP}}$ filter \\
    \texttt{eGrp} & error on the $G_{\mathrm{RP}}$ mag \\
    \texttt{PI} & parallax [mas] \\
    \texttt{ePI} & error on parallax [mas] \\
    \texttt{muRa} & proper motion on ra [mas/yr] \\
    \texttt{emuRa} & error on muRa [mas/yr] \\
    \texttt{muDec} & proper motion on dec [mas/yr] \\
    \texttt{emuDec} & error on muDec [mas/yr] \\
    \texttt{P} & membership probability \\
    \texttt{member} & 1 for sources that passed our members selection \\
    \hline
    \end{tabularx}
    \label{tab:catalog}
\end{table}

With this paper we publicly release a catalogue of the region described in Section \ref{sec:data}.
The catalogue contains photometry of 210\,907 sources in the \textit{Sloan}-like filters
$u,\,g$ and $i$ and {\it Gaia} EDR3 photometry and astrometry for the sources that are present also in the {\it Gaia} catalogue.
The sources that are not detected by {\it Gaia} have the column \texttt{gid} set to zero. The value $999$ is a placeholder for missing values (e.g., a source that has not been detected in the $u$ filter has $u=999$).
The columns of the catalogue are described in Table \ref{tab:catalog}.
The catalogue and the stacked image are available at the following url: \url{https://web.oapd.inaf.it/bedin/files/PAPERs_eMATERIALs/M37_ugiSchmidt/}.

\section{Conclusions}

We reduced and analysed Schmidt images of the open cluster M37 combining our photometric catalogue with {\it Gaia} EDR3. 
We developed software tools to correct for the geometric distortion exploiting the {\it Gaia} reference system in the case of data from wide-field imagers collected with large-dithers.
The set of routines that we developed for this specific instrument will be applied also to other wide-field imagers.

We have astrometrically and photometrically identified seven isolated WDs as candidate cluster members. We obtained follow-up low resolution spectra for one of them, confirming it as a hot, hydrogen-deficient (pre-) WD. This star was previously identified as the likely central star of a faint PN \citep{chornay2020}. Further follow-up spectroscopy is needed to univocally confirm it as one rare example of CSPN belonging to a Galactic open cluster. By means of spectral energy distribution analysis, we suggest that four out of seven WD candidates are likely or very likely cluster members. Follow-up spectroscopy is also needed to confirm their cluster membership, eventually joining the already numerous family of degenerate stars found in this rich open cluster \citep{2015ApJ...807...90C,cummings2016}.

We also publicly released a catalogue of 210\,907 sources in a $\sim 2\times2$\,deg$^2$ region centred on M37, complementing the already existing data from IGAPS.

\section*{Acknowledgements}
Based on observations collected at Schmidt and Copernicus telescopes
(Asiago, Italy) of INAF.
Partly based on observations collected with DOLORES@TNG under Director Discretionary Time program A44DDT3.
This work has also made use of data from the European Space Agency (ESA) mission
{\it Gaia} (\url{https://www.cosmos.esa.int/gaia}), processed by the {\it Gaia}
Data Processing and Analysis Consortium (DPAC,
\url{https://www.cosmos.esa.int/web/gaia/dpac/consortium}). Funding for the DPAC
has been provided by national institutions, in particular the institutions
participating in the {\it Gaia} Multilateral Agreement.
MG, LRB, DN, LS and AV acknowledge support by MIUR under PRIN program \#2017Z2HSMF and by PRIN-INAF\,2019 under program \#10-Bedin.
RR has received funding from the postdoctoral fellowship program Beatriu de Pin\'os, funded by the Secretary of Universities and Research (Government of Catalonia) and by the Horizon 2020 programme of research and innovation of the European Union under the Maria Sk\l{}odowska-Curie grant agreement No 801370.
This research makes use of public auxiliary data provided by ESA/Gaia/DPAC/CU5 and prepared by Carine Babusiaux.
This research used the facilities of the Italian Center for Astronomical Archive (IA2) operated by INAF at the Astronomical Observatory of Trieste.
This research has made use of the HASH PN database at \url{hashpn.space}.

\section*{Data Availability}

All Schmidt data are publicly available at the INAF archive\footnote{\url{http://archives.ia2.inaf.it/aao/}} (P.I. Bedin). The {\it Gaia} EDR3 catalogue is accessible through the {\it Gaia} archive\footnote{\url{https://gea.esac.esa.int/archive/}}.



\bibliographystyle{mnras}
\bibliography{bibliography} 



\appendix

\section{Additional figures}
In this section we show the finding charts and the best fit results for the SED analysis of six WD candidates.


\begin{figure*}
    \centering
    \begin{subfigure}{.46\textwidth}
    \centering
    \includegraphics[width=\textwidth]{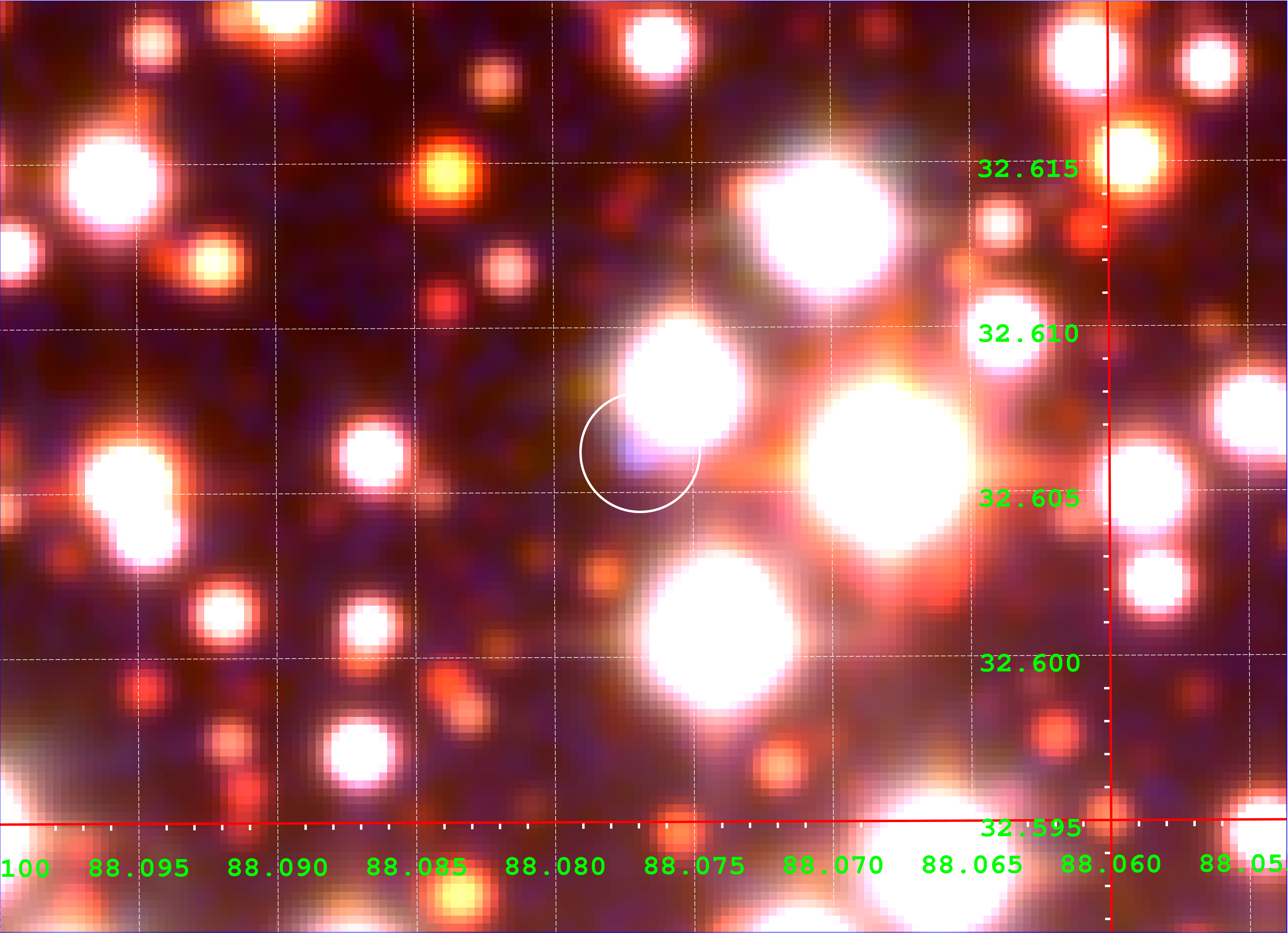}
    \caption{WD\,2}
    \label{fig:wd2}
    \end{subfigure}
    \hfill
    \begin{subfigure}{.46\textwidth}
    \centering
    \includegraphics[width=\textwidth]{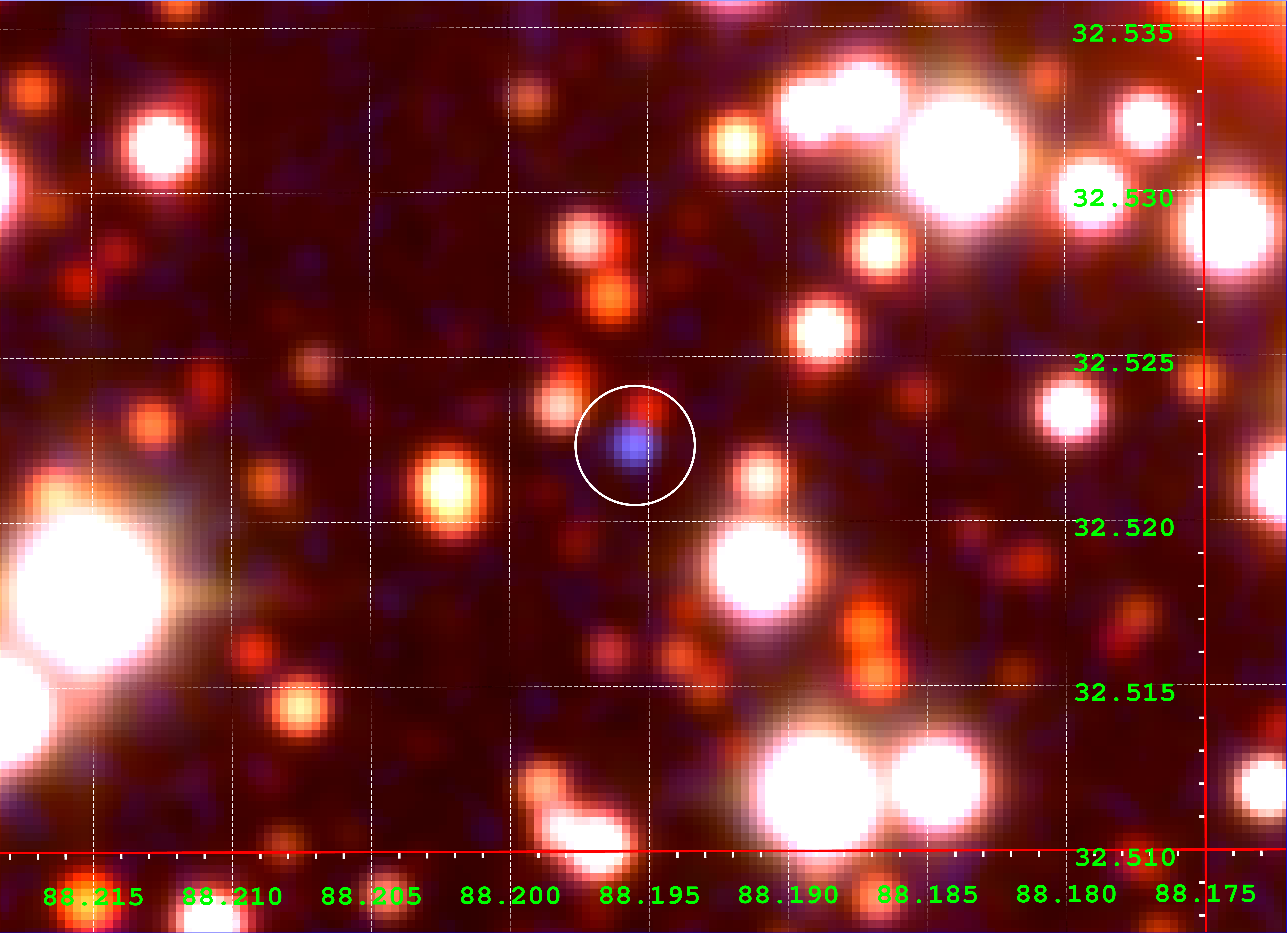}
    \caption{WD\,3}
    \label{fig:wd3}
    \end{subfigure}
    \hfill
    \begin{subfigure}{.46\textwidth}
    \centering
    \includegraphics[width=\textwidth]{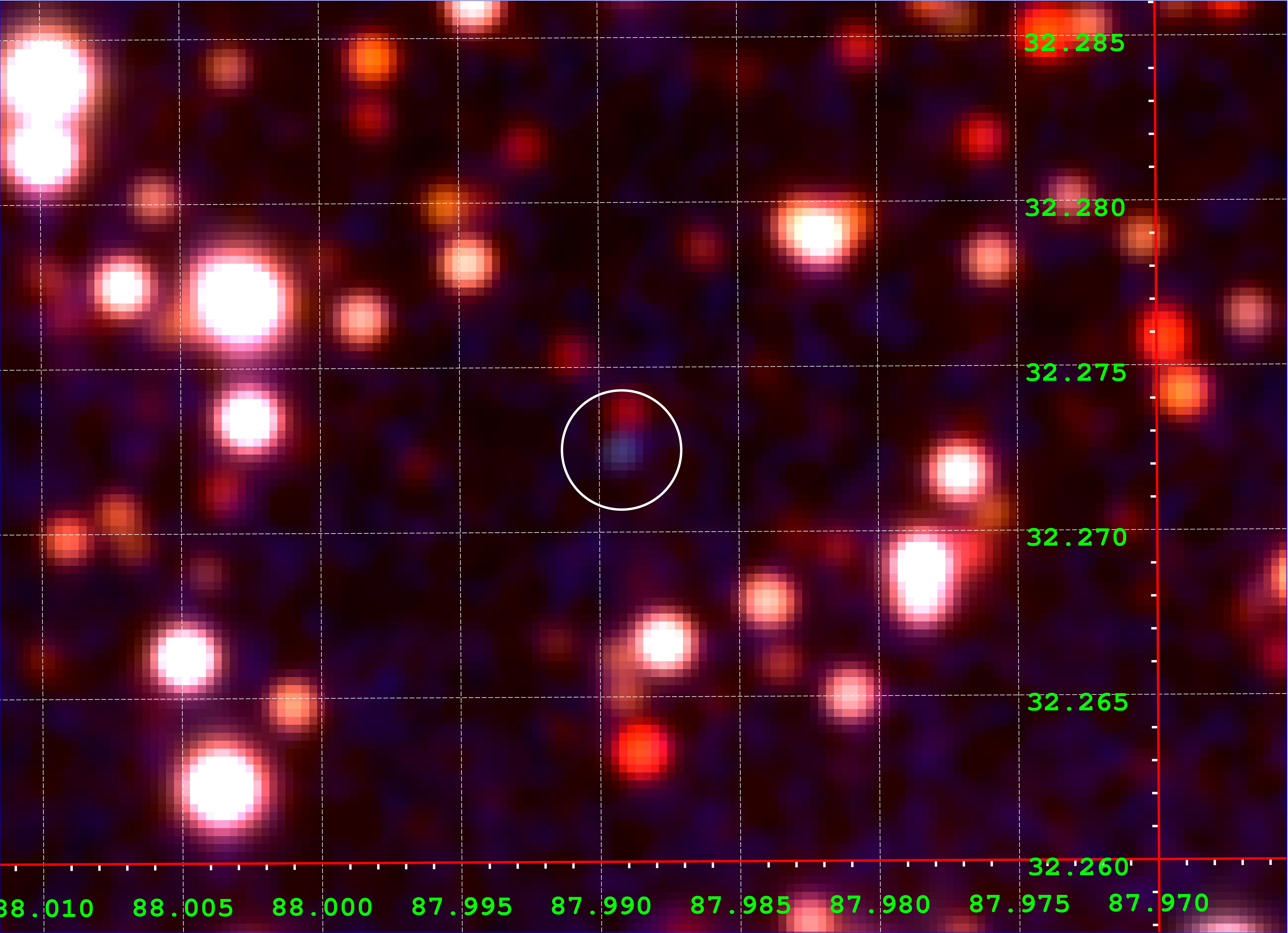}
    \caption{WD\,4}
    \label{fig:wd4}
    \end{subfigure}
    \hfill
    \begin{subfigure}{.46\textwidth}
    \centering
    \includegraphics[width=\textwidth]{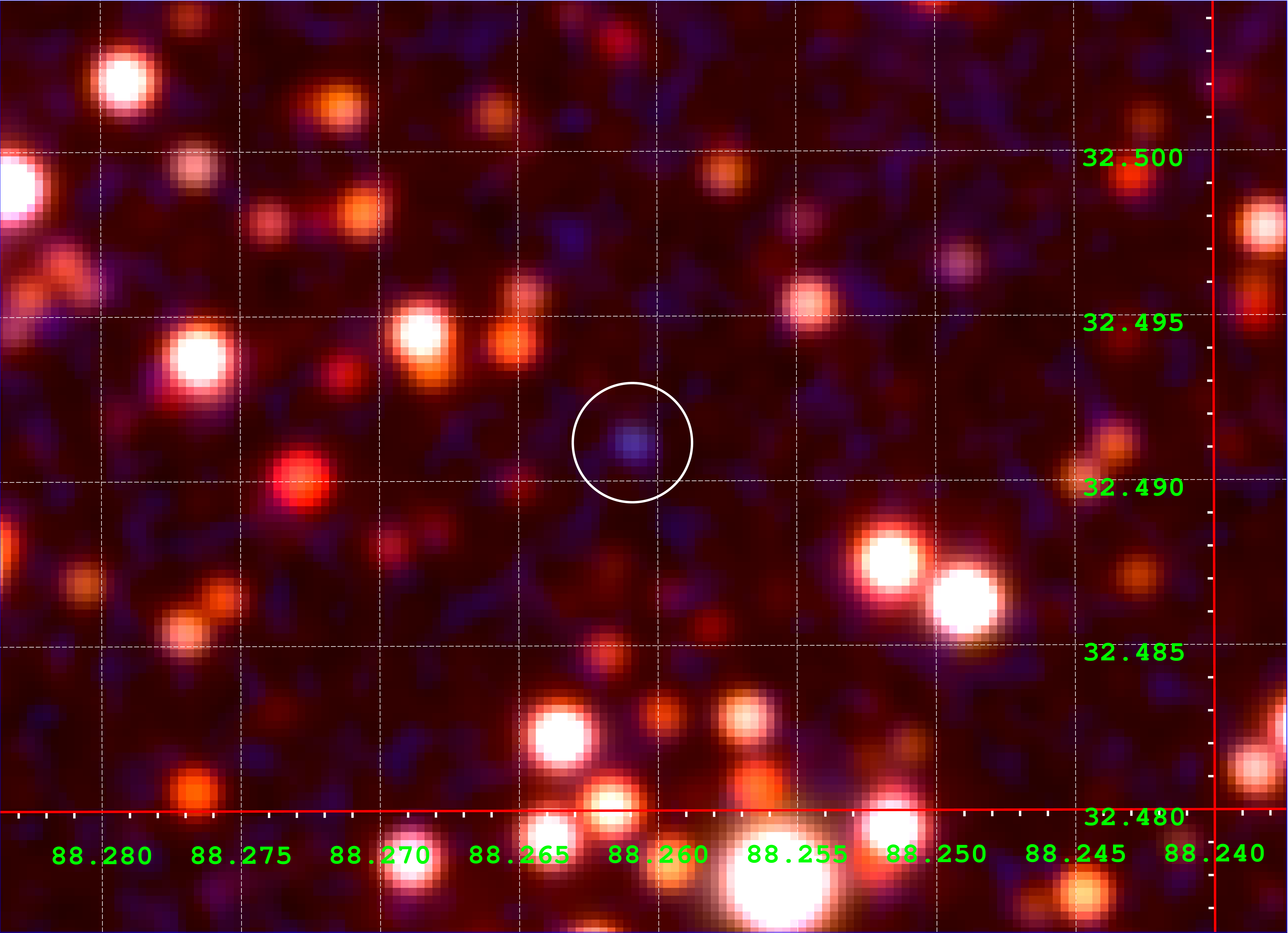}
    \caption{WD\,5}
    \label{fig:wd5}
    \end{subfigure}
    \hfill
    \begin{subfigure}{.46\textwidth}
    \centering
    \includegraphics[width=\textwidth]{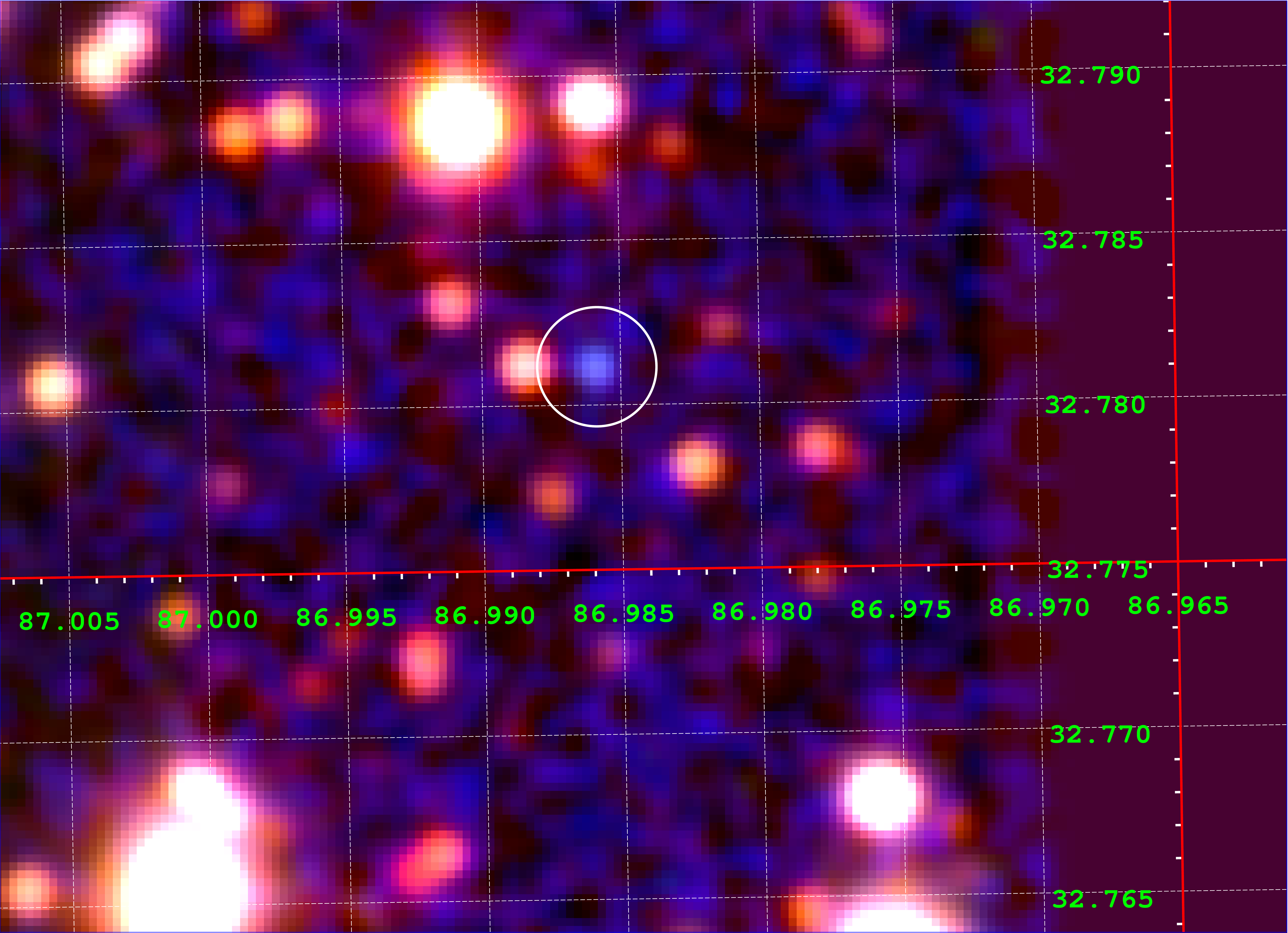}
    \caption{WD\,6}
    \label{fig:wd6}
    \end{subfigure}
    \hfill
    \begin{subfigure}{.46\textwidth}
    \centering
    \includegraphics[width=\textwidth]{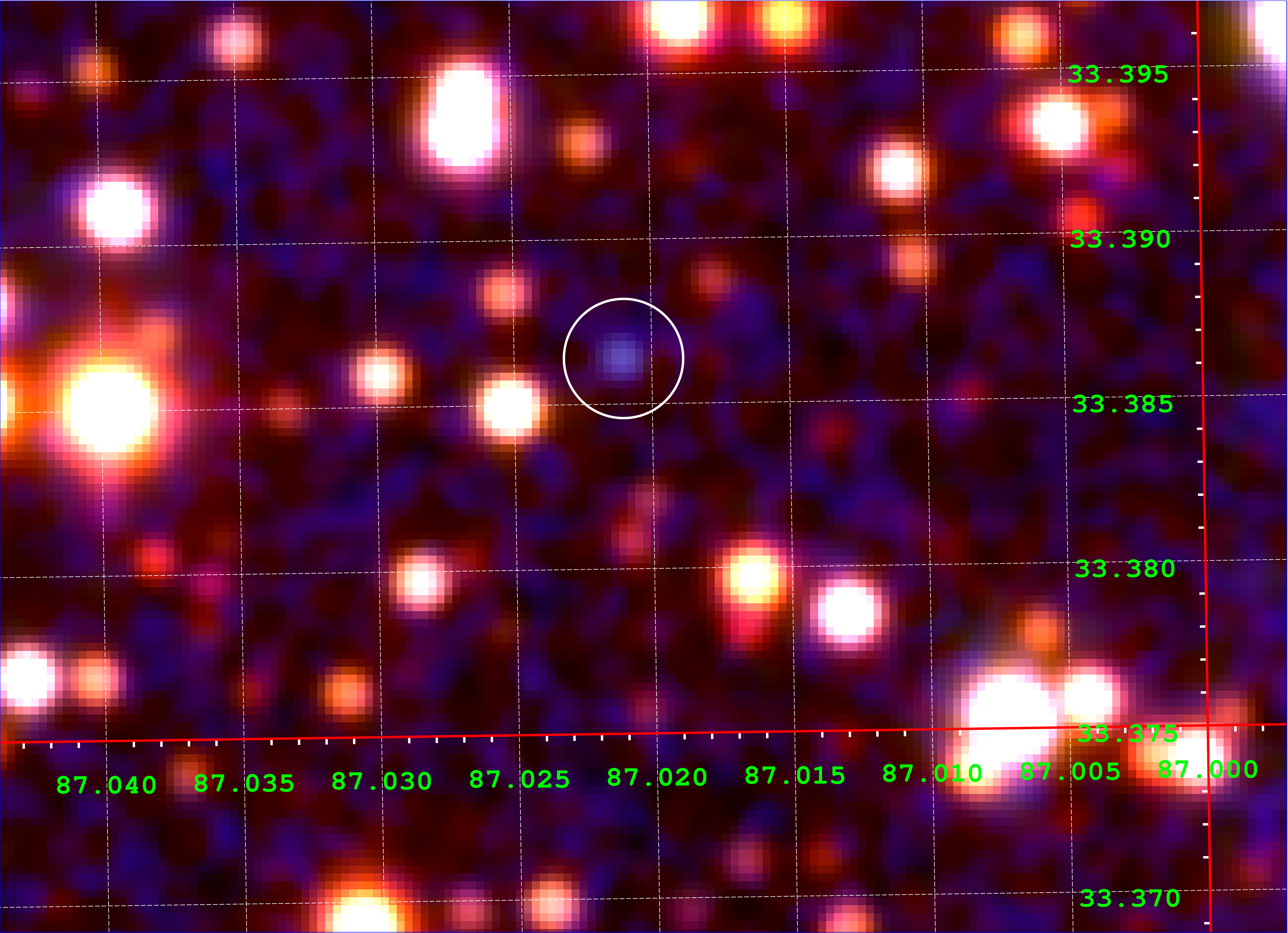}
    \caption{WD\,7}
    \label{fig:wd7}
    \end{subfigure}
    \caption{Finding charts of six WD candidates.}
    \label{fig:wdfc}
\end{figure*}

\begin{figure*}
    \centering
    \begin{subfigure}{0.47\textwidth}
    \centering
    \includegraphics[width=\textwidth]{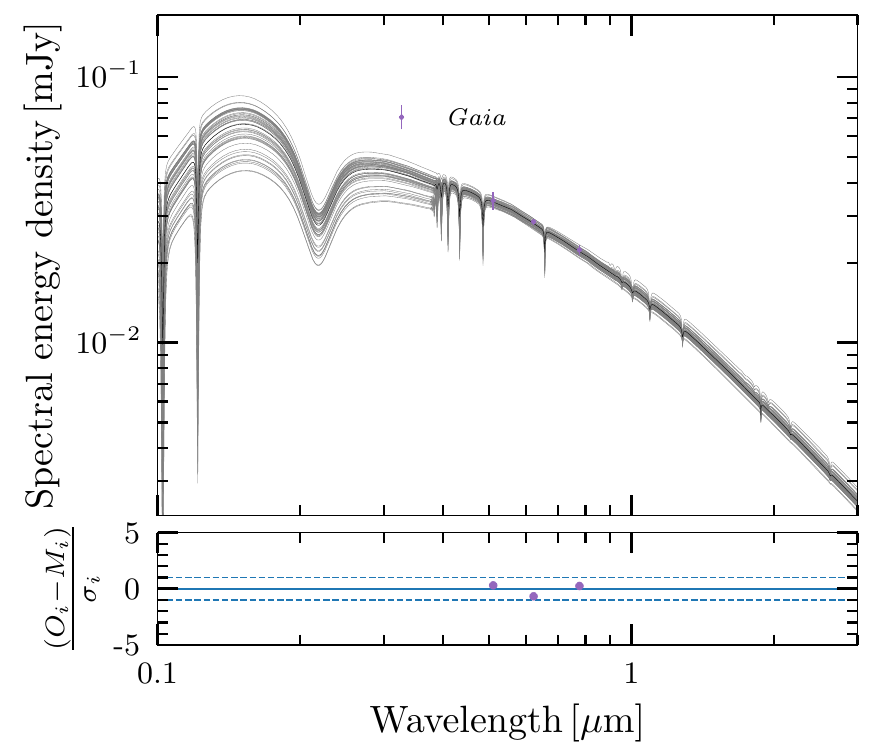}
    \caption{WD\,2}
    \label{fig:sed2}
    \end{subfigure}
    \hfill
    \begin{subfigure}{0.47\textwidth}
    \centering
    \includegraphics[width=\textwidth]{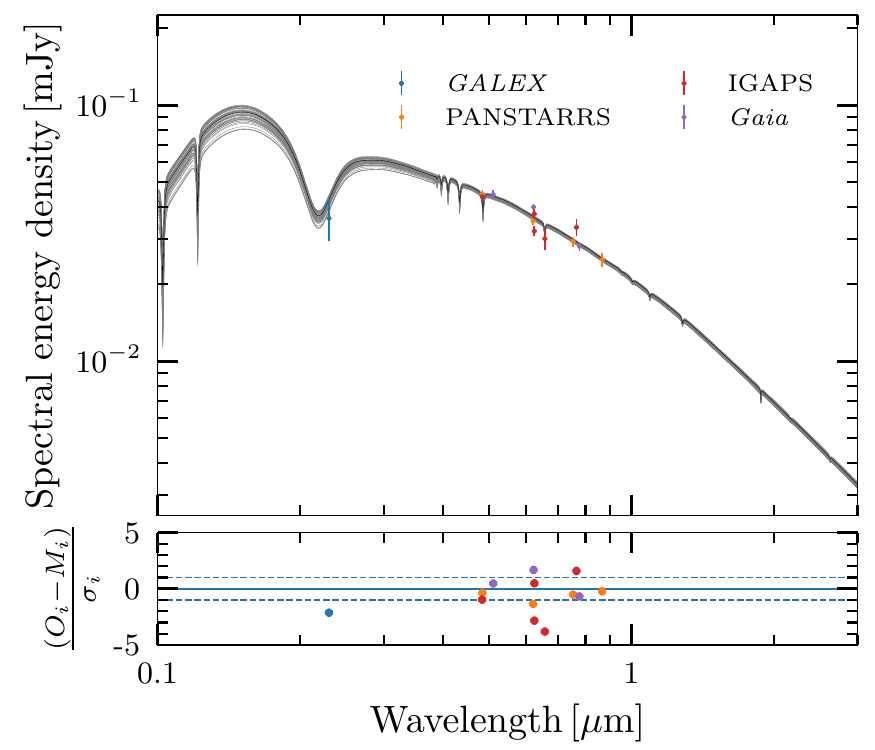}
    \caption{WD\,3}
    \label{fig:sed3}
    \end{subfigure}
    \hfill
    \begin{subfigure}{0.47\textwidth}
    \centering
    \includegraphics[width=\textwidth]{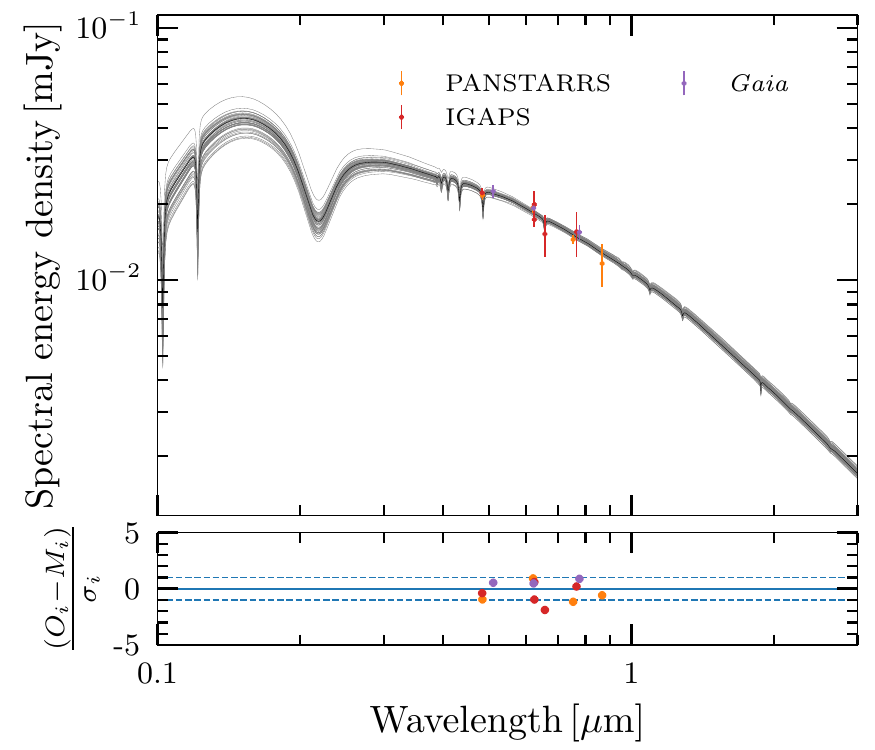}
    \caption{WD\,4}
    \label{fig:sed4}
    \end{subfigure}
    \hfill
    \begin{subfigure}{0.47\textwidth}
    \centering
    \includegraphics[width=\textwidth]{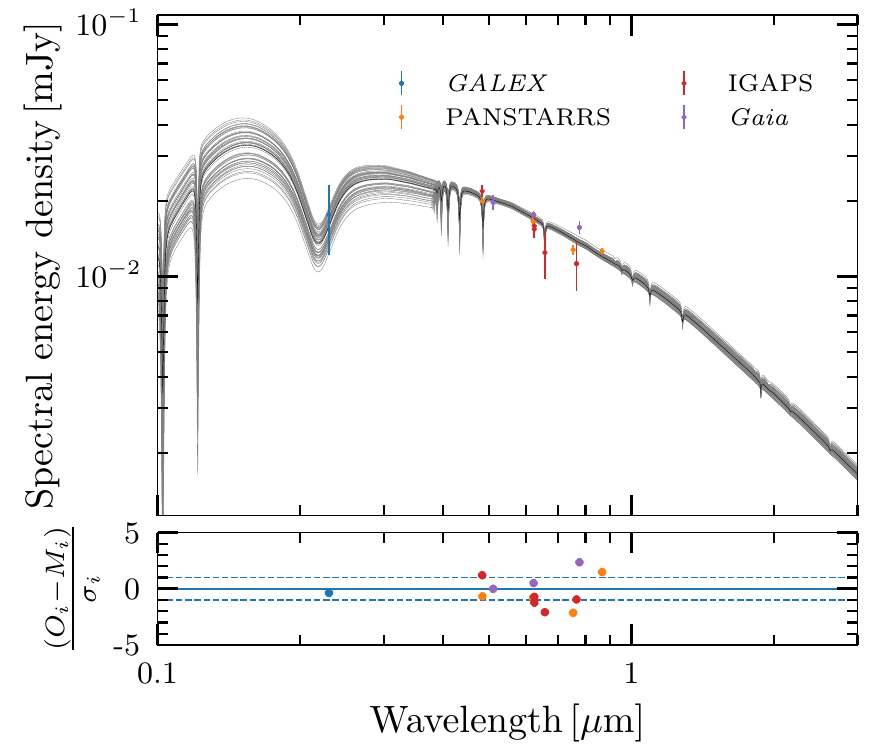}
    \caption{WD\,5}
    \label{fig:sed5}
    \end{subfigure}
    \hfill
    \begin{subfigure}{0.47\textwidth}
    \centering
    \includegraphics[width=\textwidth]{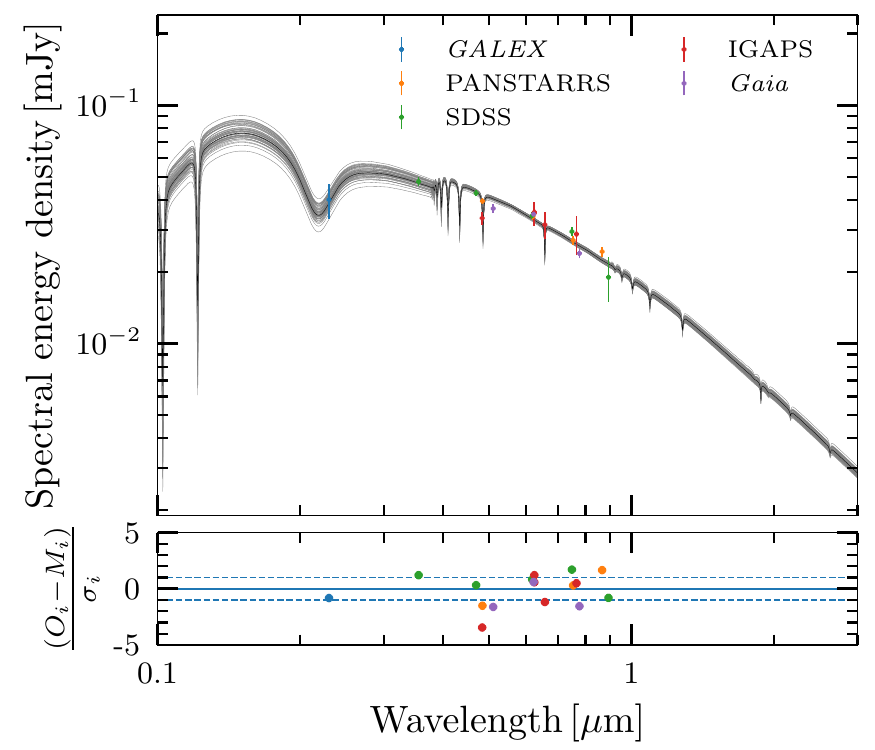}
    \caption{WD\,6}
    \label{fig:sed6}
    \end{subfigure}
    \hfill
    \begin{subfigure}{0.47\textwidth}
    \centering
    \includegraphics[width=\textwidth]{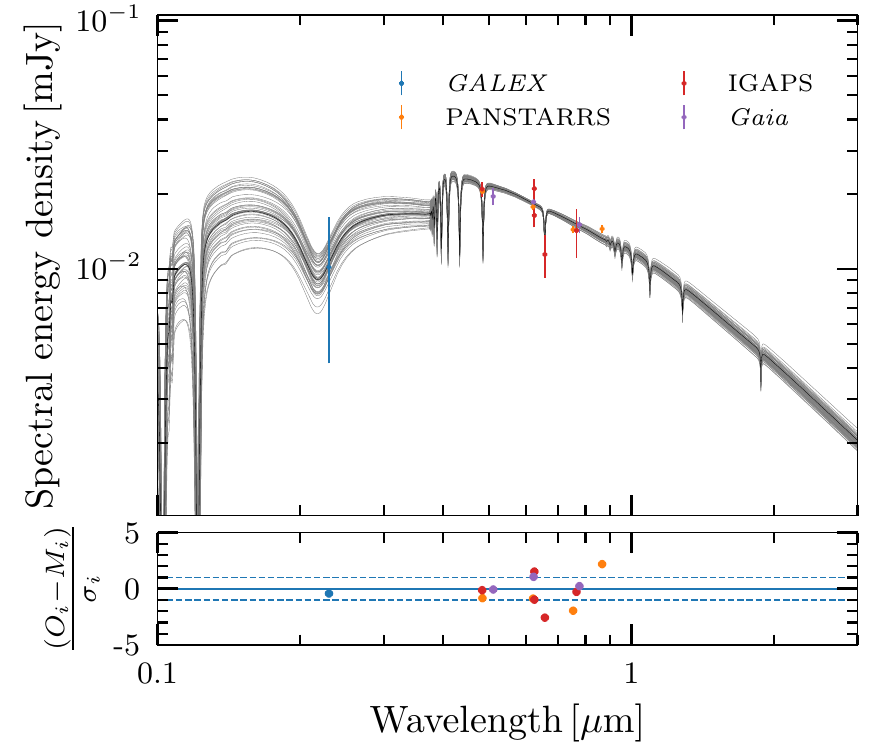}
    \caption{WD\,7}
    \label{fig:sed7}
    \end{subfigure}
    \caption{Best fit results for the SED analysis of six WD candidates. In each panel, the upper plot shows the best-fit model (black) and the models corresponding to the uncertainty range (grey). The bottom plots represent the residuals among the used photometry and the best-fit model.}
    \label{fig:seds}
\end{figure*}


\bsp	
\label{lastpage}
\end{document}